\documentclass[twocolumn,trackchanges]{aastex631}
\usepackage{lipsum}
\usepackage{amssymb}
\usepackage{float}

\begin{document}

\newcommand\lsim{\mathrel{\rlap{\lower4pt\hbox{\hskip1pt$\sim$}}
\raise1pt\hbox{$<$}}}
\newcommand\gsim{\mathrel{\rlap{\lower4pt\hbox{\hskip1pt$\sim$}}
\raise1pt\hbox{$>$}}}

\newcommand{\CS}[1]{{\color{teal} CS: #1}}
\newcommand{\sn}[1]{{\color{blue} SN: #1}}
\newcommand{\KEB}[1]{{\color{cyan} KEB: #1}}
\newcommand{\AR}[1]{{\color{red} AR: #1}}
\newcommand{\BH}[1]{{\color{orange} BH: #1}}
\newcommand{\IA}[1]{{\color{green} IA: #1}}
\newcommand{\AS}[1]{{\color{purple} AS: #1}}

\newcommand{\GAIA}{\textit{Gaia}}
\newcommand{\COSMIC}{\texttt{COSMIC }}
\newcommand{\norm}[1]{\lvert #1 \rvert}


\title{\Large Once a Triple, Not Always a Triple: The Evolution of Hierarchical Triples that Yield Merged Inner Binaries}
\shorttitle{Once a Triple, Not Always a Triple}
\shortauthors{Shariat et al.}

\author[0000-0003-1247-9349]{Cheyanne Shariat}
\affiliation{Department of Astronomy, California Institute of Technology, 1200 East California Boulevard, Pasadena, CA 91125, USA}
\affiliation{Department of Physics and Astronomy, University of California, Los Angeles, Los Angeles, CA 90095, USA}

\author[0000-0002-9802-9279]{Smadar Naoz}
\affiliation{Department of Physics and Astronomy, University of California, Los Angeles, Los Angeles, CA 90095, USA}
\affiliation{Mani L. Bhaumik Institute for Theoretical Physics, University of California, Los Angeles, Los Angeles, CA 90095, USA }

\author[0000-0002-6871-1752]{Kareem El-Badry}
\affiliation{Department of Astronomy, California Institute of Technology, 1200 East California Boulevard, Pasadena, CA 91125, USA}

\author[0000-0003-4189-9668]{Antonio C. Rodriguez}
\affiliation{Department of Astronomy, California Institute of Technology, 1200 East California Boulevard, Pasadena, CA 91125, USA}

\author[0000-0001-7840-3502]{Bradley M.S. Hansen}
\affiliation{Department of Physics and Astronomy, University of California, Los Angeles, Los Angeles, CA 90095, USA}
\affiliation{Mani L. Bhaumik Institute for Theoretical Physics, University of California, Los Angeles, Los Angeles, CA 90095, USA }

\author[0000-0002-9751-2664]{Isabel Angelo}
\affiliation{Department of Physics and Astronomy, University of California, Los Angeles, Los Angeles, CA 90095, USA}

\author[0000-0001-8220-0548]{Alexander P. Stephan}
\affiliation{Department of Physics and Astronomy
Vanderbilt University, Nashville, TN 37235, USA}

\keywords{binaries: close, merged – binaries: general – white dwarfs – stars: general, triples – stars: kinematics and dynamics}

\correspondingauthor{Cheyanne Shariat}
\email{cshariat@caltech.edu}

\begin{abstract}
More than half of all main-sequence (MS) stars have one or more companions, and many of those with initial masses $<8$~M$_\odot$ are born in hierarchical triples. These systems feature two stars in a close orbit (the inner binary) while a tertiary star orbits them on a wider orbit (the outer binary). In hierarchical triples, three-body dynamics combined with stellar evolution drives interactions and, in many cases, merges the inner binary entirely to create a renovated `Post-Merger Binary' (PMB). By leveraging dynamical simulations and tracking binary interactions, we explore the outcomes of merged triples and investigate whether PMBs preserve signatures of their three-body history. Our findings indicate that in $26$-$54\%$ of wide double WD binaries ($s\gsim 100$~au), the more massive white dwarf (WD) is a merger product, implying that these DWD binaries were previously triples. Overall, we estimate that $44\pm14\%$ of observed wide DWDs originated in triple star systems and thereby have rich dynamical histories. We also examine MS+MS and MS+Red Giant mergers manifesting as Blue Straggler stars (BSSs). These PMBs have orbital configurations and ages similar to most observed BSS binaries.  While the triple+merger formation channel can explain the observed chemical abundances, moderate eccentricities, and companion masses in BSS binaries, it likely only accounts for $\sim20$-$25\%$ of BSSs. Meanwhile, we predict that the majority of observed single BSSs formed as collisions in triples and harbor long-period ($>10$~yr) companions.
Furthermore, both BSS binaries and DWDs exhibit signatures of WD birth kicks.


\end{abstract}

\section{Introduction} \label{sec:introduction}

Observational and theoretical studies clearly show that most main-sequence (MS) stars form and evolve with one or more stellar companion \citep[e.g.,][see latter for a review]{Raghavan2010, Tokovinin14, Moe17, Offner23}. For multiple star systems, there are numerous possible processes that govern the early stages of formation. Fragmentation models suggest that most multiple star systems form via the gravitational collapse of over-dense regions that arise in cores, filaments, or massive accretion disks \citep[e.g.,][]{Larson72, Adams89, Inutsuka92, Goodman93, Kratter08, Krumholz09, Meyer18}. Dynamical capture, evolution, and migration models supply even more channels for the formation of multiple systems \citep[e.g.,][]{Ostriker94, Bate97, Ostriker99, Tohline02, Moeckel07, Bate12, Lee19, Cournoyer21}.

Naturally, and from stability arguments, the protostars that form in these regions tend towards hierarchical configurations \citep{Duchene13}. In triple systems, for example, hierarchy manifests with two stars residing in a close orbit (the inner binary) while a tertiary star orbits around them on a wider orbit (the outer binary). In dense open and globular clusters, where dynamical interactions are more frequent, hierarchical triples are commonplace and help shape the stellar population \citep[e.g.,][]{Latham07, Geller09, Leigh11, Geller12, Milone12, Leigh13, Geller13}. Stellar triples are also abundant in the galactic field \citep[e.g.,][]{Raghavan2010, Tokovinin14}. For example, $40\%$ of short-period binaries with $0.5-1.5$~M$_\odot$ companions have an additional companion \citep{Tokovinin1997, Tokovinin14}. Amongst contact binaries, a similar $42\%$ are part of triples \citep{Pribulla06}. Further studies have demonstrated the abundance of triple systems by observed eclipsing binaries \citep{Rappaport13, Conroy14} and stellar multiplicity studies \citep{Tokovinin1997, Eggleton07, Moe17}. Recently, \citet{Shariat23} performed a detailed investigation on the evolution of solar-type stars in hierarchical triples from their birth to their fate as white dwarfs (WDs). By comparing their dynamical simulations and \textit{Gaia} observations, they predicted that $30\%$-$40\%$ of solar-type stars in the galactic neighborhood were formed in hierarchical triples. 

During the evolution of hierarchical triples, the distant tertiary star will exchange angular momentum with the inner binary in a secular fashion, causing eccentricity-inclination oscillations: a process known as the eccentric Kozai-Lidov (EKL) mechanism \citep[][see the latter for a review]{vonZeipel1910, Kozai1962, Lidov1962, Naoz2016}. Large eccentricities induced in the inner binary from EKL cycles lead triples to interact (i.e., transfer mass) far more often than they would as isolated binaries \citep[e.g.,][]{Naoz2014, Stephan16, Cheng+19, Toonen20, Shariat23}\footnote{Other hierarchical perturbations, such as flyby stars, can also excite high eccentricities in wide systems, leading to interactions \citep[e.g.,][]{Kaib2014,Michaely2016,Michaely2019b,Michaely2020,Michaely2021}}. This makes triples an especially interesting channel to study binary exotica such as Blue Straggler stars \citep[BSSs;][]{Geller13}, cataclysmic variables \citep[CVs;][]{Nelemans01, Knigge11, Pala17, Knigge22}, and Post-Common Envelope Binaries \citep[e.g.,][]{Toonen13, Zorotovic14, Hernandez22,Yamaguchi24b}.

Beyond just mass transfer, interactions in the inner binary catalyzed by a distant companion can cause the inner binary to merge entirely \citep[e.g.,][]{Antonini17, Grishin22, Trani22, Preece22, Rajamuthukumar23, Shariat23, Dorozsmai24}. In such cases, the triple will become a binary where one star is a merger remnant and the other is the tertiary from the previous triple. We refer to these triples-turned-binaries as Post-Merger Binaries (PMBs).

In this study, we explore the signatures of the evolution of PMBs from their triple birth to their binary fate. We investigate whether PMBs contain observable signatures that allude to their triple origin. For instance, the remnant star can be bluer and brighter than the main-sequence turnoff \citep[e.g.,][]{Sandage53}, have unusual chemical abundances \citep[e.g.,][]{Hansen16}, have rapid or inclined rotation \citep[e.g.,][]{Qureshi+18,Leiner19, Stephan+20,Toonen2022}, or contain two components with apparently discrepant ages \citep[i.e., a non-coeval binary;][]{Heintz22}.

We organize our study of triples-turned-binaries as follows. In Section \ref{sec:methodology}, we outline the simulations that were used to evolve the triples and binaries for Myr to Gyr timescales. We then explore the three body merger channel to form apparently discrepant wide DWDs (Section \ref{subsec:DWDs}) and BSS binaries (Section \ref{subsec:BSSs}). In Section \ref{sec:WD_MSRG}, we examine the orbital configurations of triples with merged WD+MS or WD+RG inner binaries. In Section \ref{sec:other_outcomes}, we discuss other interesting outcomes of triples with mergers and their applications to observable stellar systems. Finally, in Section \ref{sec:conclusions}, we summarize our main results and conclusions.

\section{Methodology} \label{sec:methodology}

\subsection{Physical Processes and Numerical Setup}\label{sec:num}

\citet{Shariat23} simulated $\sim4000$ hierarchical triple systems and dynamically evolved all three stars for over $10$~Gyr. Following their setup, we consider a hierarchical triple system of masses $m_1$, $m_2$, on a tight inner orbit, and $m_3$ on a wider orbit about the inner binary. The three stars have radii $R_1$, $R_2$, and $R_3$. The orbital parameters include the inner (outer) semi-major axis $a_1$ ($a_2$), eccentricity $e_1$ ($e_2$), and inclination with respect to the total angular momentum $i_1$ ($i_2$). In \citet{Shariat23}, a robust sample ($\gsim50\%$) of the three-body simulations that assumed a Kroupa Initial Mass Function \citep[IMF,][]{Kroupa2001} resulted in a merged inner binary, turning the triple in a binary. In this study, we focus on the subset of these systems that lived on as `Post-Merger Binaries' (PMBs).

In the simulations, \citet{Shariat23} solve the equations of motion for the hierarchical triple up to the octupole level of approximation \citep[see][for the full set of equations]{Naoz2016}. They also include general relativistic precession up to the 1st post-Newtonian order for the inner and outer orbit \citep[e.g.,][]{Naoz2013GR}. These precession approximations are sufficient to describe the dynamics within the mass ratios and scales studied here \citep[e.g.,][]{Naoz2013GR,Lim20,Kuntz22}. Equilibrium tides are adopted for both inner binary members, following \citet{Hut80}, \citet{Eggleton98}, and \citet{Kiseleva98}. This model includes rotational precession, tidal precession, and tidal dissipation  \citep{Naoz2014}. For stars on the giant branch, the code switches to radiative tides following \citep{Stephan18, Rose+19}.

Stellar evolution plays a crucial role in the evolution of triples \citep[e.g.,][]{Naoz2016,Naoz+16,Toonen2016,Toonen20,Toonen2022,Cheng+19,Hamers22,Stegmann22,Kummer23, Shariat23}. For example, the mass loss associated with the Asymptotic Giant Branch (AGB) phase can re-trigger the EKL mechanism \citep{PK12,Shappee13,Michaely2014,Naoz+16,Stephan16,Stephan17}. We thus follow the post-main sequence evolution of stars using the Single Stellar Evolution (\texttt{SSE}) code \citep{SSE}.  See \citet{Naoz2016,Naoz+16,Stephan16,Stephan17,Stephan18,Stephan19,Stephan21} and \citet{Angelo22} for a detailed description of the triple with stellar evolution code. If the inner binary experiences mass transfer or becomes tidally locked, the evolution of the inner binary is carried out using \texttt{COSMIC} \citep{COSMIC}, a suite for the Binary Stellar Evolution (\texttt{BSE}) code \citep{BSE} with additional modifications. We outline our general procedure for transferring between the three-body code and \texttt{COSMIC} in Section \ref{subsec:Evolving_Beyond_Merger}. In that section, we also give an example evolution of a triple system that merged (Figure \ref{fig:Single_triple_merger}). For a more descriptive outline of implementing \texttt{COSMIC}, refer to \citet{Shariat23}.

The three-body simulations from \citet{Shariat23} included two different models with slightly different initial ansatzes. Both models use a Kroupa IMF \citep{Kroupa2001} from $1-8$~M$_\odot$, but use different methods of sampling the initial inner and outer orbital periods. The first model chooses the inner and outer orbital periods independently from a log-normal period distribution with a mean of $\log(4.8/d)$ and a standard deviation of $\log(2.3/d)$ 
\citep{DM91}. Then, the larger value is taken to the period of the outer orbit, and the smaller is taken as the period of the inner orbit. As the last step, we keep only systems that pass the hierarchy and stability criterion. Hierarchy is determined by the inequality \citep{Naoz2016}:
\begin{equation}\label{eq:eps_crit}
    \epsilon = \frac{a_1}{a_2}\frac{e_2}{1-e_2^2} < 0.1 \ .
\end{equation} 
The stability criterion we use is \citep{MA2001}:
\begin{equation}\label{eq:MA_stability_crit}
    \frac{a_2}{a_1}>2.8 \left(1+\frac{m_3}{m_1+m_2}\right)^\frac{2}{5} \frac{(1+e_2)^\frac{2}{5}}{(1-e_2)^\frac{6}{5}} \left(1-\frac{0.3i}{180^\circ}\right) \ .
\end{equation}

We label this set of runs as `IB' (Independent Binary) following \citet{Shariat23} because both the inner and outer period distributions are re-sampled independently. Therefore, this model effectively tests formation models where the inner and outer periods have no correlation. This could reflect independent fragmentation models and dynamical capture models \citep[e.g.,][]{Ostriker94, Meyer18}. 

\begin{figure}
\includegraphics[width=1.0
\columnwidth]{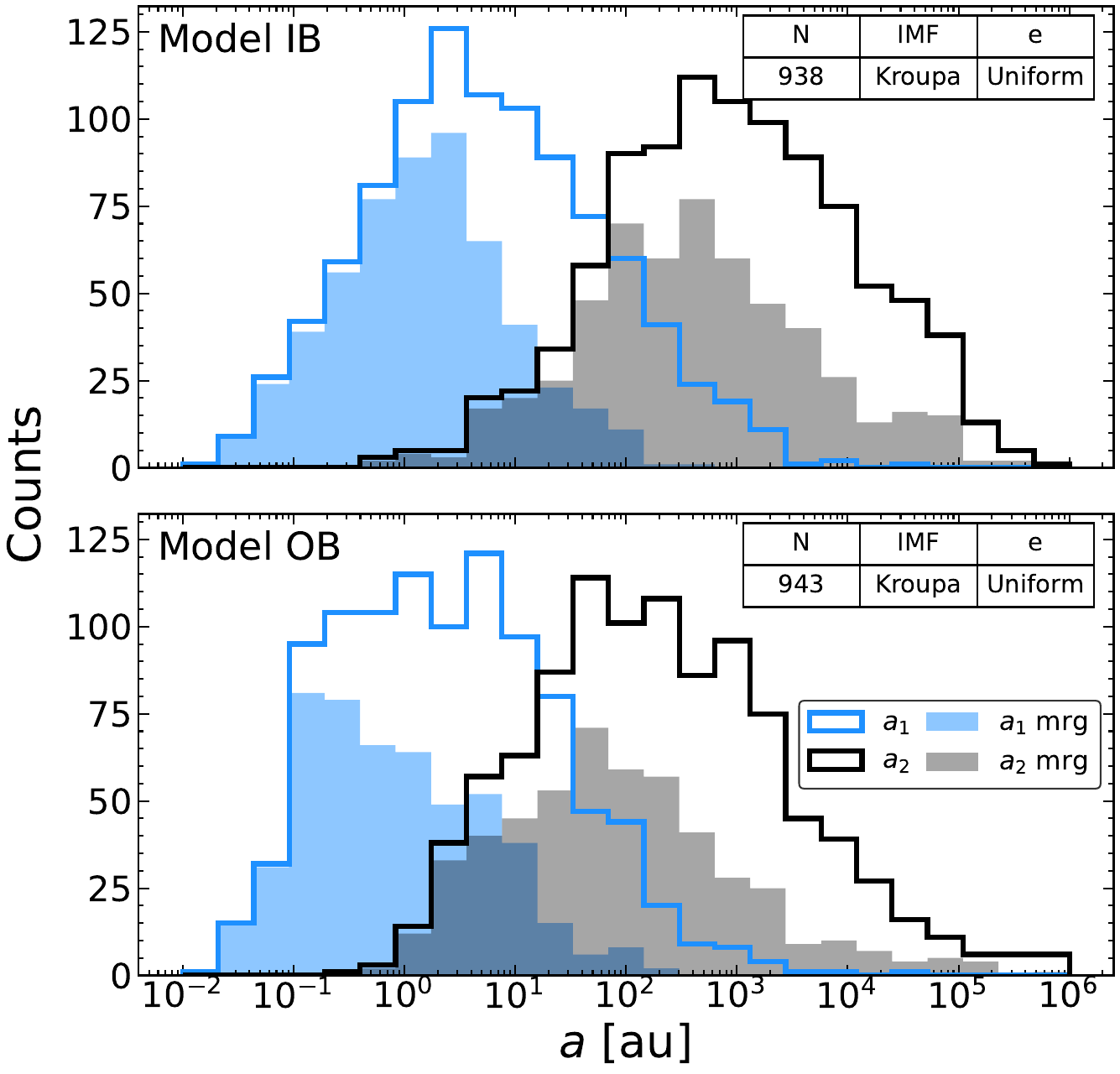}
\caption{Initial Conditions of the triple-star models. The solid line represents semi-major axes of all triples while the shaded regions shows them for triples that experienced a merged inner binary during the three-body evolution.
Both the inner and outer semi-major axes distributions were sampled from \citet{DM91}, but the bottom panel had $a_2$ sampled from \citet{DM91}, and for that fixed value of $a_2$, $a_1$ was sampled until a stable combination was formed. While in the top panel, both $a_1$ and $a_2$ were sampled independently from the distribution (see Section \ref{sec:methodology}). In Model IB, $59\%$ ($N=553$) of all of the triples experienced a merged inner binary, and in Model OB, this fraction was $54\%$ ($N=509$) \citep{Shariat23}. In the tables, we show the total number of initial triples, their IMF, and initial eccentricities. The initial inclinations were chose to be isotropic for all triples. }\label{fig:ICs} 
\end{figure}

\citet{Shariat23} also devises a second channel for sampling the orbital periods from the \citet{DM91} distribution. In this model, they sample the period of the outer orbit from the \citet{DM91} distribution, and for that chosen orbit, resample the period of the inner binary until a stable system is formed. If the sampled outer period is too small to physically allow a stable inner orbit, we resample the outer orbit. In this second channel for generating initial periods, a hierarchy of formation is assumed: the outer orbit formed first and limited the allowed (stable) configurations of the inner orbit. We label this set of runs as `OB' (Outer Binary) to denote that the outer binary has governance. Overall, the different sampling methods means that triples in the OB model have closer inner binaries more often (Figure \ref{fig:ICs}).

The initial semi-major axis distribution for Model IB is displayed in the top panel of Figure \ref{fig:ICs}. We also show the initial $a_1$ (blue) and $a_2$ (black) distribution for the triples that led to a merged inner binary. The $a_1$ and $a_2$ distributions for Model IB closely resemble two Gaussians, whereas in the Model OB $a_2$ distribution is effectively Gaussian, while the $a_1$ distribution is a left-biased Gaussian constrained by the $a_2$ value. Overall, most inner binaries with close initial separations ($a_1\lsim 10$~au) experienced a merger. Furthermore, the inner binaries with closer tertiaries, which would induce stronger EKL effects, were more likely to merge. In Figure \ref{fig:ICs}, we also show the total number of triples (N) for each model at $t=0$, the IMF information, and the eccentricity distribution (assumed uniform for both). The initial inner and outer inclinations are chosen from an isotropic distribution.

\vspace{-0.85cm}
\begin{deluxetable*}{lccccccccccc}
\tablecaption{Inner Binary Mergers\label{tab:merger_types_models}}
\tablehead{
\colhead{} &
\colhead{N} &
\colhead{N Merged} &
\colhead{N Formed} &
\colhead{N} &
\colhead{N Double} &
\multicolumn{6}{c}{Merger Type} \\
\cline{7-12}
\colhead{} & 
\colhead{Triples} & 
\colhead{Total} & 
\colhead{PMB} & 
\colhead{Disrupted} & 
\colhead{Merger} & 
\colhead{MSMS} &  
\colhead{RGMS} & 
\colhead{RGRG} &
\colhead{WDMS} &
\colhead{WDRG} &
\colhead{WDWD} 
}
\startdata
Model~\textit{IB} & 938 & 540 & 415 & 125 & 9 & 79 & 175 & 8 & 68 & 34 & 50 \\
Model~\textit{OB} & 943 & 473 & 395 & 78 & 27  & 106 & 133 & 13 & 92 & 24 & 27
\enddata
\end{deluxetable*}

In the IB Model, $59\%$ ($N=553$) of all of the triples experienced a merged inner binary, and in the OB model, this fraction is $54\%$ ($N=509$) \citep{Shariat23}. Across both models, $65$ had destroyed inner binaries (i.e., no remnant star) and another
$119$ triples were unbound or disrupted by the merger event. Also, $19$ PMBs were unable to be modeled in \texttt{COSMIC} beyond their merger age. This leaves behind $395$ ($415$) survived PMBs in the IB (OB) model. The inner binaries that merged to produce PMBs are diverse, including mergers between MS, Red Giant (RG), and WD stars. 

In Table \ref{tab:merger_types_models}, we show the abundance of the different outcomes from each model, including the various types of binary mergers. In this table, we include the number of total completed triple simulations (`N Total') along with the number of those that had merged inner binaries (`N Merged Total'). `N Formed PMB' refers to the number of triples that had merged inner binaries and survived as PMBs. `N Disrupted' counts the times when the inner binary merger disrupted the triple. This includes cases where the merger did not leave behind a surviving star and cases where the merger event unbound the tertiary from the system. `N Double Merger' is the number of times where the inner binary and post-merger outer binary merged (discussed in Section \ref{subsec:double_mergers}). In the last six columns, we show the distribution for the different types of mergers among only the merged triples that survive as bound PMBs. The OB model had closer initial inner and outer binary separations, which led to a greater fraction of early MSMS mergers and double mergers. On the other hand, the IB model had a greater fraction of mergers that occurred later in the triple evolution, leading to more mergers involving RG and WD stars. 

For both the IB and OB models, we also calculate the effect that kicks during WD formation would have on the orbits. Different sources have found support for WD kicks, and we briefly outline some here. Based on observations of the globular clusters and Monte Carlo modeling, WD kicks of a few km/s prolong the core contraction of clusters, and this resolves the discrepancy between theory and observations \citep{Fregeau09}. Moreover, WD birth kicks of ($\sim 0.75$~km/s) resolve the discrepancy between Binary Population Synthesis models and the observed separation distribution of WD binaries from Gaia \citep{EB18}. \citet{Shariat23} also found that accounting for an agnostic $\sim 0.75$~km/s kick was essential to reproduce the observed separation distribution of WD triples. More recently, measurements of WDMS binaries with current separations of a few au show that most binaries have eccentricities above $\sim0.15$ and up to $0.8$ \citep{Shahaf24}. The separation of these systems suggests that they underwent mass transfer, though they are likely too distant to have experienced a common-envelope phase \citep[e.g.][]{Lagos22, Yamaguchi24}. Nevertheless, during mass transfer, the binaries should have presumably circularized through tides. WD kicks are one proposed mechanism to explain such binaries, though other mechanisms have not been ruled out \citep[see][and references therein]{Shahaf24}.

Considering the strong possibility of WD kicks, we add the new models that assume an agnostic and modest kick during WD formation for all systems of the IB and OB models. The minor kick changes the eccentricity and semi-major axis of the outer orbit, generally by pumping the eccentricity and widening the orbit; we follow \citet{Lu2019} to calculate the new orbital parameters. For the complete protocol of applying a WD kick to the three-body evolution, refer to \citet{Shariat23}.

\subsection{Evolving Beyond Merger}\label{subsec:Evolving_Beyond_Merger}

Of the entire sample of triple star systems from \citet{Shariat23}, $\sim 50\%$ had an inner binary that merged. These represented the systems where the inner binary (1) began to interact in the triple code and (2) after being put into \texttt{COSMIC}, were found to coalesce into one star. We put the binary into \texttt{COSMIC} to follow the moments leading up to the inner binary merging. In most cases, the merger leaves behind one star, creating a newly-formed binary with the tertiary from the triple ($m_3$). 

During the time that the inner binary was undergoing its interaction, we also followed the evolution of $m_3$ on these short timescales (often $\sim$~Myr) to track any changes in mass or radius. If there were any changes to the tertiary that would change the orbital configuration during this time, they are applied. We calculate the new semi-major axis and eccentricity of the post-merger binary based on (1) the previous triple's orbital parameters and (2) how long the inner binary interacted. If the interaction of the inner binary was larger than the outer orbital period (between the inner binary and the tertiary), then we assumed an adiabatic mass loss, and therefore, the eccentricity remained the same, and the semi-major axis expanded accordingly. If the mass loss occurred on times smaller than the orbital period, we applied a `kick' prescription to model the mass loss as being effectively instantaneous. In this case, the new eccentricity and semi-major axis were calculated following \citet{Lu2019}. For more details on our procedure to calculate the orbital parameters of the PMB, see \citet{Shariat23}.

Now that the triple has become a PMB and its orbital configuration is identified, we again put the newly-formed PMB into \texttt{COSMIC} and evolve until completion (often a few Gyr for most of the following analysis). In many cases, the triple and binary evolution is highly involved and leads to an interesting array of final binaries that were previously triples. We outline one example evolution in Figure \ref{fig:Single_triple_merger}. We caution that both at this stage and in the interacting inner binary stage, {\tt COSMIC} sets the eccentricity to zero. Thus, in all of our predictions, we underestimate the eccentricity of the short-period binaries, including PMBs.



\subsubsection{Example Evolution of a Post-Merger Binary}\label{subsubsec:example_evolution}

\begin{figure*}
\centering
\includegraphics[width=1.0
\textwidth]{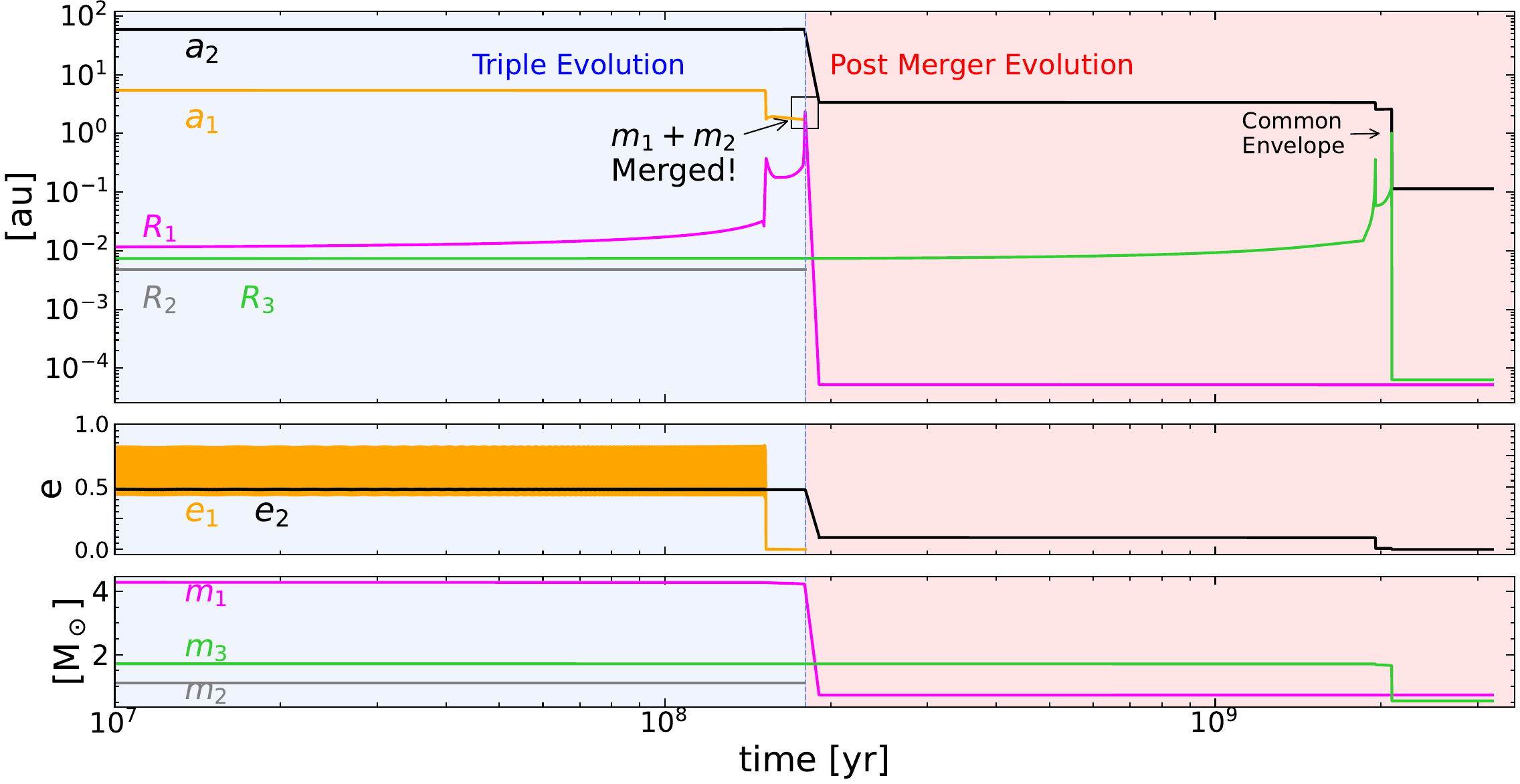}
\caption{Example evolution of a triple that experienced a merged inner binary and became a post-merger binary. In the top panel, we track the evolution of the inner and outer semi-major axes ($a_1$ and $a_2$) and the radii for all three stars in the triple ($R_1, R_2, R_3$). In the middle panel, we track the eccentricity of the inner ($e_1$) and outer ($e_2$) orbits. In the bottom panel, we show the evolution of the mass for all three stars  ($m_1, m_2, m_3$). The blue shaded region (labeled `Triple Evolution') is when the system was still a triple, and the red (labeled `Post Merger Evolution') shows its evolution as a binary after the two inner stars merged at $t\sim 189$~Myr. 
Initially, $m_1=4.3$~M$_\odot$, $m_2=1.1$~M$_\odot$, $m_3=1.7$~M$_\odot$, $a_1 = 5.4$~au, $a_2 = 59.2$~au, $e_1 = 0.45$, and $e_2 = 0.47$. The more massive $m_1$ evolves first onto the Red Giant Branch and then the Asymptotic Giant Branch. During this time, it begins contact with $m_2$, which strips most of its envelope. Quickly, $m_2$ spirals into $m_1$ and leaves behind a C/O white dwarf with $0.72$~M$_\odot$. This white dwarf is now in a binary orbit with the tertiary $m_3=1.7$~M$_\odot$ In the `Post Merger Evolution' (red) region, the binary evolves steadily until $m_3$ starts moving off the main sequence and expands at $t=1.35$~Gyr. Here, it undergoes common envelope evolution with the primary WD at $1.52$~Gyr, at which point the orbit has shrunk significantly, and $m_3$ becomes a C/O WD. The end result is a double WD system in a $12$~day period.
}.\label{fig:Single_triple_merger} 
\end{figure*}

In Figure \ref{fig:Single_triple_merger}, we present an example of the evolution of 
a stellar triple that has 
$m_1=4.3$~M$_\odot$, $m_2=1.1$~M$_\odot$, $m_3=1.7$~M$_\odot$, $a_1 = 5.4$~au, $a_2 = 59.2$~au, $e_1 = 0.45$, and $e_2 = 0.47$. The tertiary here excites only moderate eccentricity excitation (middle panel). After $\sim 189$~Myr, the more massive star ($m_1$) in the inner binary transitions from core Helium burning and approaches the first Asymptotic Giant Branch (AGB). Here, it inflates into Roche crossing, where it shrinks and circularizes the inner orbit and begins a sequence of events where the binary goes in and out of contact evolution. In the $1$~Myr before the stars merged entirely, $m_2$ strips most of $m_1$'s envelope during the first common envelope evolution, which removes $\sim3$~M$_\odot$ from the binary and leaves $m_1$ as a $0.93$~M$_\odot$ Naked Helium star in the Hertzsprung Gap. During the final moments of the binary's life, $m_2$ spirals into $m_1$, expelling $\sim 1 M_\odot$ of mass and pushing the remnant star to the AGB. After this merger event settles, the new primary survives as an inflated AGB star (the result of $m_1+m_2$ merging) and has an MS companion ($m_3$ from the triple). Tides from the inflated AGB star quickly shrink and circularize the new binary's orbit to $a = 3.4$~au. However, the enriched AGB star quickly has its mass ejected and forms a $0.72$~M$_\odot$ WD at $t=190$~Myr.

We are left with a new binary where the primary is a $0.72$~M$_\odot$ WD and the tertiary is $m_3=1.7$~M$_\odot$ MS star. Now, entering the `Post Merger' (red) region of the plot, simulated with \texttt{COSMIC}, the binary evolves steadily until $m_3$ starts to move off the main sequence. As $m_3$ expands at $t=1.35$~Gyr it undergoes common envelope evolution with the merged product, now the primary WD. The mass ejection persists until $1.52$~Gyr, at which point the orbit has shrunk significantly, and $m_3$ is now a C/O WD, making a double WD system with a period of $12$ days. This stage utilizes {\tt COSMIC}, which assumes efficient tides and thus sets the eccentricity to zero.

The remaining binary is an example of a post-merger binary that ended its eventful life as a circular DWD binary with an orbital period of $12$ days, a primary mass of $0.72$~M$_\odot$, and a secondary mass of $0.55$~M$_\odot$. Throughout the evolution, the system could have been observed as a (1) Blue Straggler Binary system during the RGMS merger \citep[e.g.,][]{Leiner19}, (2) cataclysmic variable-like binary during the second mass transfer episode, and (3) DWD with anomalous cooling ages, as observed in \citet{Heintz22}. Note that for $\sim800$~Myr during the post-merger evolution, the system was a WD+MS binary separated by a few au. Observed $\sim1$~au WD+MS binaries have a deficit of massive WD components ($\gsim 0.8$~M$_\odot$), which may hint at a lack of merger products \citep{Hallakoun23}. We also find that the shortest WDMS binary in which the WD is a result of a previous WD+WD merger is $\sim10$~au, supporting that close WDMS systems are unlikely to have undergone a previous merger.

\section{Double White Dwarfs}\label{subsec:DWDs}
\subsection{Observations of Double White Dwarfs}\label{subsubsec:DWDs_obs}
Stars with initial masses less than $\sim8-10 M_\odot$ become white dwarfs. As a result, WDs are extremely common and represent the endpoint for $97\%$ of stars in the Galaxy \citep{Fontaine01}. After isolated single stars become white dwarfs, they no longer undergo nuclear fusion and, therefore, strictly cool over time. The cooling age of a WD is a function of its birth mass, with higher initial masses leading to longer cooling times and lower initial masses leading to shorter cooling times for single WDs. Therefore, accurate cooling models \citep[e.g.,][]{Bergeron95} allow single WDs to serve as precise age indicators.

\citet{Heintz22} studied the ages of WDs in wide DWD binaries from the Gaia Early Data Release 3 (eDR3). The age of WDs in this sample was determined using a Monte-Carlo SED-fitting approach from broad-band photometry. Recently, \citet{Heintz24} did a follow-up study where they re-derived the age estimated from spectroscopic data and found consistent values with the photometric evidence. Interestingly, they identify $283$ DWDs with anomalous age measurements, representing $21\%-36\%$ of their sample. In these DWDs, the more massive WD possesses a shorter cooling age than its less massive companion with over $3\sigma$ confidence. This discrepancy was observed with both photometric and spectroscopic age estimates \citep{Heintz22, Heintz24}, and it suggests that the less massive WD was born from a more massive main-sequence progenitor. This observation is not consistent with a monotonically increasing initial-final mass relation (IFMR), which leads to new interpretations.

One potential explanation for these peculiar observations is that the more massive WD in the binary is the product of a prior merger event, suggesting that the DWD system was originally a triple. One example of a WD being formed from two merging stars is shown in the $m_1$+$m_2$ merger in Figure \ref{fig:Single_triple_merger} and described in Section \ref{subsubsec:example_evolution}. Here, an MS star merged with an AGB star, forming an already-evolved C/O WD. This merger remnant WD would presumably have a shorter cooling age than would be estimated from the isolated stellar evolution model for a WD of its mass.

\citet{Shariat23} found that upwards of $30\%$ of stars in the solar neighborhood were born as triples, with over half of them experiencing a merger. This proposes a formation channel for the $21\%-36\%$ of DWDs that may potentially have merger remnants \citep{Heintz22}. A $20\%$ merger fraction is also consistent with binary population synthesis models \citep{Temmink20} and the multiplicity rate of WD progenitor stars \citep{Moe17}, making the potential for a prior merger a great possibility. 

Furthermore, \citet{Hallakoun23} recently discovered an order-of-magnitude deficit of massive WD companions in $\sim 1$~au binaries with F/M-dwarfs. At such separations, the binaries were likely to have undergone a phase of stable mass transfer and are unlikely to harbor a merger remnant because the stars are close. Since the massive WDs ($>0.8$~M$_\odot$) are missing in this sample, a significant fraction of massive WDs in the field may be merger products \citep{Temmink20, Hallakoun23}. Given the likelihood that a fraction field WDs are the result of previous mergers, we investigate whether the three-body formation channel can dynamically support a high merger rate of WDs in the galactic field (see Section \ref{subsubsec:DWDs_obs_to_sim}).

\subsection{Comparing Observed Double White Dwarfs to Theoretical Simulations }\label{subsubsec:DWDs_obs_to_sim}

\begin{figure*}
\includegraphics[width=1.0
\textwidth]{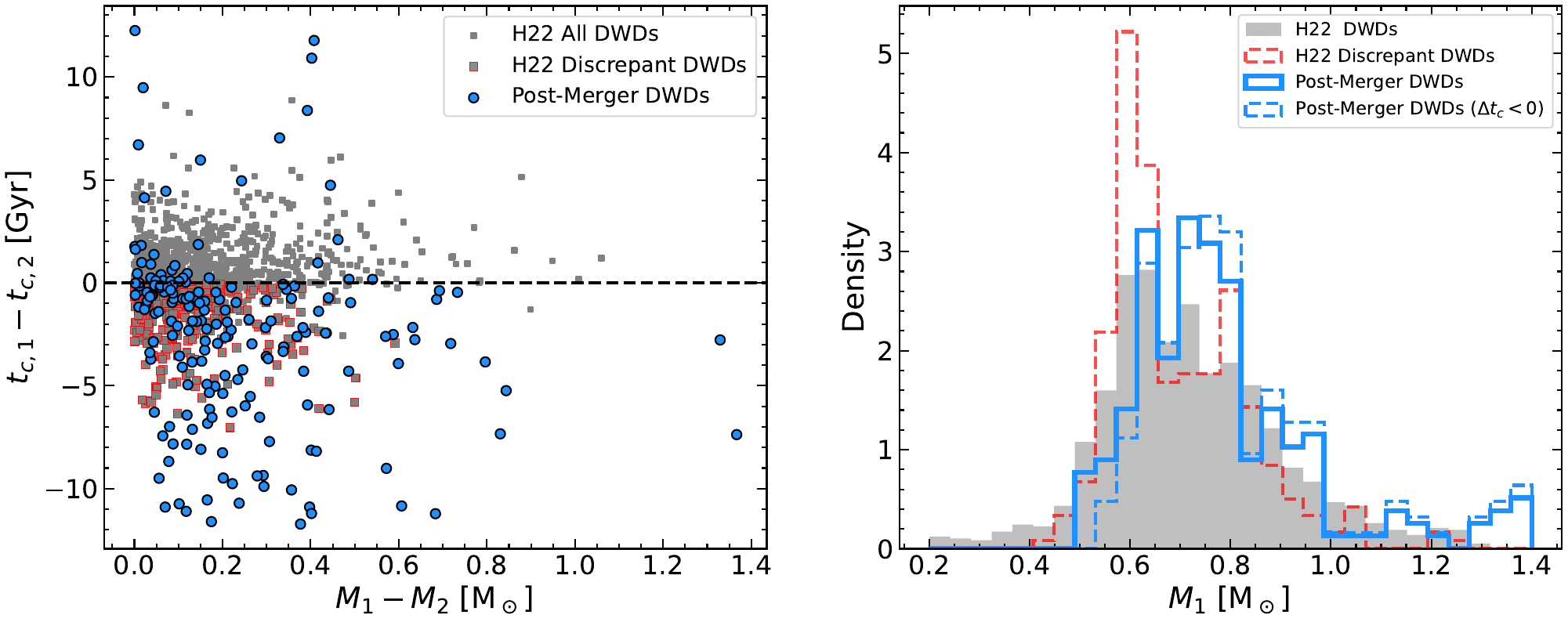}
\caption{The cooling ages and mass of post-merged binaries that became DWDs (`Post-Merger DWDs') compared to observed DWDs with apparently discrepant ages from \citet{Heintz22}.
\textbf{Left:} The difference in the cooling age between the more-massive DWD (subscript `1') and the less-massive WD (subscript `2') compared to their difference in mass. In gray, we show the observed systems, where those with a red outline are the discrepant DWDs. In blue, we show our simulated systems that experienced a merger and were DWDs at $t=13.7$~Gyr.
\textbf{Right:} The normalized distribution of the primary WD mass in observed wide DWDs (gray), observed discrepant DWDs (red dashed), the simulated post-merger DWDs (blue), and the simulated discrepant post-merger DWDs (blue dashed). The histograms are normalized such that the integrated area under the curve is 1.
}\label{fig:DWD_cool_mass} 
\end{figure*}

\subsubsection{Cooling Ages and Masses }\label{subsubsec:DWDs_coolages}
To calculate the total age of an observed WD, \citet{Heintz22} summed the best-fit progenitor age (its main-sequence lifetime) and the best-fit WD age based on its cooling track.  To calculate the best-fit progenitor age, they mapped the current mass of the WD to a progenitor's ZAMS mass using a shifted MESA IFMR \citep{Fields16}. From the ZAMS mass, one can then calculate the total main sequence lifetime, assuming single stellar evolution. If there were a merger during the star's evolution, this age calculation method would give the wrong answer, leading to a discrepancy amongst the ages of stars in the binary DWDs. Since the study focuses on wide systems \citep[$s\gsim100$~au,][]{Heintz22}, the probability of previous mass transfer between the two stars is unlikely, making the merger channel a potentially robust explanation for the observed discrepancies. Here,
we seek to understand whether the PMBs that became DWDs in our theoretical simulations also possess similarly discrepant cooling ages and therefore, support the merger channel.

In Figure \ref{fig:DWD_cool_mass} we plot the cooling ages and masses of all of our post-merger DWDs after they evolved for $13.7$~Gyr (blue). The total time of evolution does affect the difference in WD cooling ages, though it does govern how large the magnitude of the difference can be. We chose a conservative value that is likely higher than the true age of observed systems, which leads to larger cooling age differences in our simulations (though most of our DWDs formed in $1$ to $10$ Gyr). In both plots, we also show the values for the observed DWDs (gray) and the discrepant subset of observed DWDs (red outline). The masses and MS/WD lifetimes of the stars were calculated using \texttt{COSMIC}. Note that the oldest and coolest WDs would be too faint to be observed with \textit{Gaia}, and therefore are missing in this sample.

Similar to Figure 7 in \citet{Heintz22}, the left plot of Figure \ref{fig:DWD_cool_mass} shows the cooling age difference as a function of the mass difference between the more-massive and less-massive WDs (subscript `1' and `2', respectively.). Of the $163$ post-merger DWDs in our sample, $80\%$ exhibit discrepant cooling ages. In these systems, the merger history of the primary WD made the cooling age smaller than its less massive companion. To check with uncertainties, we take the median cooling age uncertainty from the systems in \citet{Heintz22} and find that it does not change the results. \citet{Heintz22} found that $21\%-43\%$ of all wide DWDs exhibit such discrepancies in their cooling ages. Based on our simulated results, we can assume that this only accounts for $80\%$ of the observed post-merger DWDs. Based on the aforementioned consistency, we attribute the anomalous ages of most observed systems to be a sign that their primary WD is a merger product. This would suggest that about $26\%-54\%$ of wide DWDs contain merger products. The presence of a merged WD today suggests that all of these systems were previously triples, like hierarchical triples. In this case, their three-body history provides more dynamical freedom compared to isolated binary evolution, which can explain their high frequency of mergers in the prior inner binary \citep[e.g.,][]{Toonen20, Shariat23}. A triple fraction of $40\pm14\%$ is consistent with \citet{Shariat23}, who predict that $\gsim30\%$ of solar-type stars were born in triples. 

The right plot of Figure \ref{fig:DWD_cool_mass} shows the distribution of the primary WD mass in the DWD, normalized so that the area under the curve is unity. We display all observed DWDs in gray, the discrepant ones in red, the simulated post-merger DWDs in blue, and the discrepant simulated DWDs in dashed blue. Compared to all DWDs, the simulated post-merger DWDs have more massive primary WDs, which is similar to the discrepant sample. The observed discrepant sample sharply peaks at $0.6$~M$_\odot$, similar to spectroscopic WD mass catalogs \citep[e.g.,][]{Liebert05, Tremblay16}. Our models peak at a similar mass and also contain a peak at slightly larger WD masses ($0.7$-$0.8$~M$_\odot$).  The sharp maximum around $0.6$~M$_\odot$ is an attribute of the WD IFMR and the IMF. Namely, most IFMRs predict that MS progenitors with masses between $1.5-2.0$~M$_\odot$ become $0.6$~M$_\odot$ WDs \citep{Fields16}, and most MS stars that are now WDs had initial masses in this range. This may indicate that observed WDs in this mass range originate from single stellar evolution or from a pre-WD merger that produced a main-sequence star in the $1.5-2.0$~M$_\odot$ mass range \citep[e.g.,][]{Liebert05}. The observed discrepant systems show a smaller peak around $M_{\text{WD}}=0.8$~M$_\odot$, which is consistent with being merger products \citep[e.g.,][]{Liebert05,Hallakoun23}. In the simulations, we find an additional local maximum at $M_{\text{WD}}>1$~M$_\odot$, which is not a feature that is observed with $Gaia$. This feature may be missing from observations because older and cooler WDs are too faint to be seen with $Gaia$. This high-mass peak may also be an artifact of the \texttt{COSMIC} models, and not be a physical manifestation of mergers \citep[see][for example]{Hansen06, Tremblay16}. Also, note that the simulated post-merger masses are heavily reliant on the post-merger mass calculation from \texttt{COSMIC} and our chosen Kroupa IMF. 
\begin{figure}[h]
\includegraphics[width=1.0
\columnwidth]{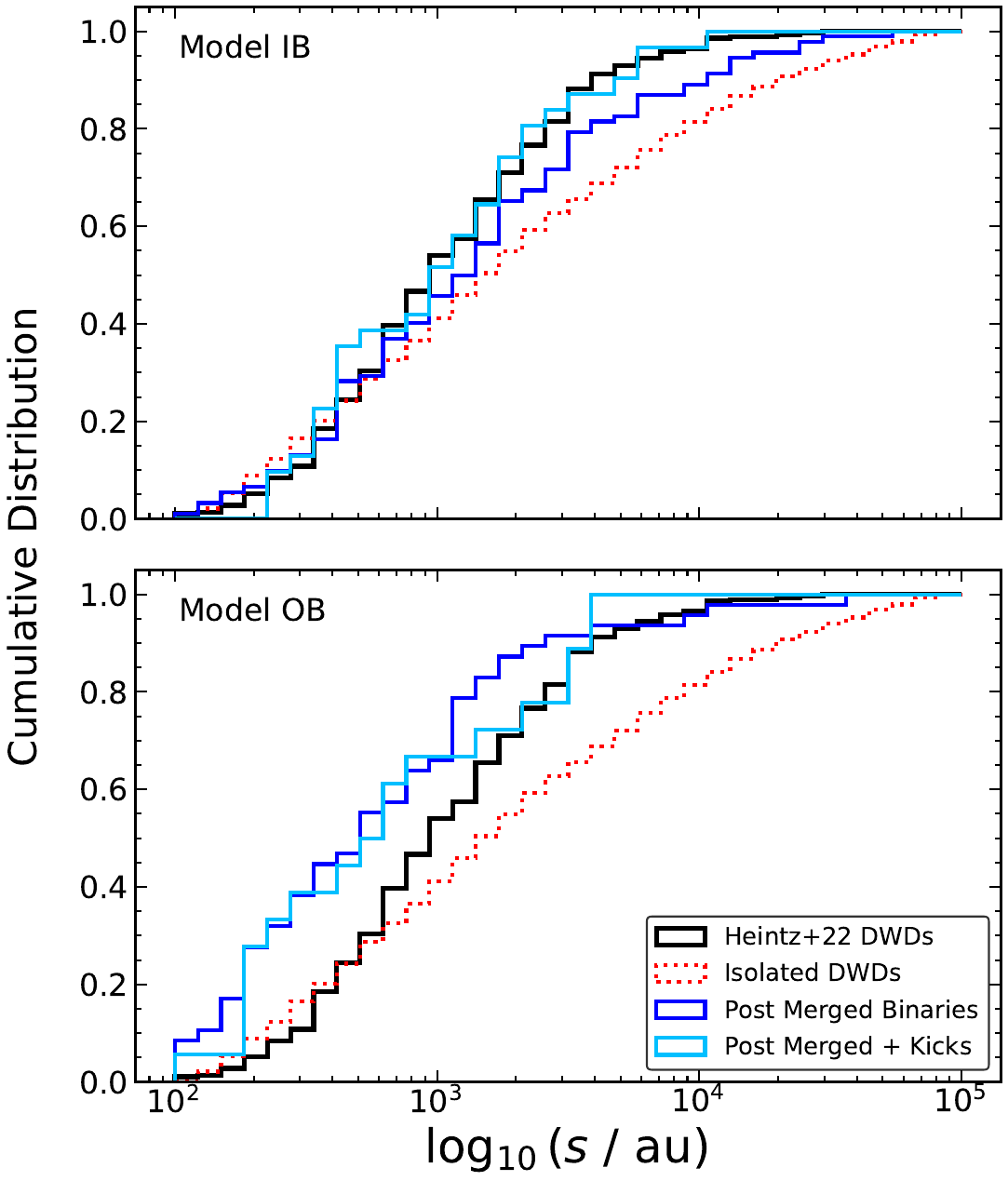}
\caption{Separation distribution of anomalous DWDs compared to simulations. We show the results of the triples that experienced an inner binary and evolved to be a WD+WD binary at t=$10$~Gyr. The dark blue curve assumed no WD birth kicks, and the light blue systems assumed WDs were born with a small kick. The dotted red line is the distribution for isolated binary evolution using \texttt{COSMIC} population synthesis. The top row shows systems from the IB model, and the bottom panels are from the OB model. The KS-test p-values associated with each of these distributions are displayed in Table \ref{table: DWD}.}\label{fig:DWD_sep_CDF} 
\end{figure}

\subsubsection{Separation Distributions }\label{subsubsec:DWDs_sep}

In Figure \ref{fig:DWD_sep_CDF}, we compare the separation distribution of our models to those observed by \citet{Heintz22}. The observed sample (black curve) contains the separations of DWDs from \citet{Heintz22} that have anomalous WD ages and have $\texttt{R\_chance\_align}<0.1$, making them confident candidates for bound DWD systems \citep[][]{EB21}. In these DWDs, the more-massive WD exhibits a shorter cooling age than the less-massive WD, with over a $3\sigma$ discrepancy \citep{Heintz22,Heintz24}. We also plot the triples in our simulations that had experienced merged inner binaries, leaving a DWD binary that was once a triple (labeled `Post Merged Binaries'). The top panel of Figure \ref{fig:DWD_sep_CDF} compares the observed distribution to the IB model, while the bottom panel compares it to the OB models. Moreover, for each panel, the light blue curve shows the same results but accounts for the effect of WD kicks. The separation is generally within a factor of unity with the semi-major axis of the orbit \citep{Dupuy11}, and for uniform eccentricities, the median conversion factor is $s=a/1.10$. Following this median relation, we plot the scaled semi-major axis of our simulations in this plot. Gaia observations are also susceptible to selection effects, and many DWDs, especially more distant ones, may be missing in the observed sample because they are unresolved. We take into account the selection effects by only including resolved systems in our simulations. Specifically, for each of our simulated DWDs, we sample $100$ possible distances based on the distances of the observed DWDs. For each random sample, we calculate whether the system would be resolved at that distance with Gaia's angular resolution of $0.43$ arcseconds \citep{Gaia_Collab}. If it is resolved for over half the randomly sampled distances, we include it in Figure \ref{fig:DWD_cool_mass} and \ref{fig:DWD_sep_CDF}. $78\%$ of all simulated systems are resolved with Gaia. Note also that many massive WDs are likely missing from the observed sample because they are smaller and, therefore, fainter.

Additionally, we simulate a population of binaries that become DWDs using \texttt{COSMIC} and compare them to all systems as well (red curve in Figure \ref{fig:DWD_sep_CDF}). We choose an initial period distribution from \citet{Raghavan2010} with uniform initial eccentricities and a Kroupa IMF. Note that the separation distribution for the binaries is often reflective of the initial separation. 

As depicted in the Figure \ref{fig:DWD_sep_CDF}, 
the IB model produces separation distributions that are consistent with the observations, more so than the OB model. Furthermore, the IB model that assumed WD kick has greater consistency with the observed DWDs. The OB systems have a more scattered separation distribution, and since most triples initially had closer tertiaries, the distribution of post-merger DWDs has an abundance of closer systems. We quantify the consistency between the various curves by performing a Kolmogorov–Smirnov (KS) statistical test with a critical p-value chosen to be $0.05$. A KS p-value below $0.05$ would suggest that the two populations likely originated from different parent distributions. Therefore, a larger KS p-value would indicate that the two distributions were unlikely to come from different parent distributions. We calculate the p-value for all of our models, including our control Binary model, against the observed DWD distribution and display the values in Table \ref{table: DWD}.
\begin{deluxetable}{lccccc}
\tablecaption{Comparison to \citet{Heintz22} DWD Distribution \label{table: DWD}}
\tablehead{
\colhead{} &
\colhead{Model} &  
\colhead{Model} & 
\colhead{Model} &
\colhead{Model} &
\colhead{Binaries} 
\\
\colhead{} &
\colhead{IB} &  
\colhead{IB+K} &
\colhead{OB} &
\colhead{OB+K} &
\colhead{}
}
\startdata
N & $101$ & $101$ & $59$ & $59$ & $1625$ \\
p-value & $0.10$ & $0.76$ & $1.7\times10^{-6}$ & $0.35$ & $4.9\times10^{-13}$ 
\enddata
\tablecomments{The p-value here is from the KS-test. `+K' denotes WD kicks. See Figure \ref{fig:DWD_sep_CDF} for the distributions.}
\end{deluxetable}

The p-values for the IB models, both without and without kicks, are greater than $0.05$, which supports the conclusion that they may come from the same parent distribution. More importantly, the p-value from isolated binary evolution models had a p-value of $\sim10^{-14}$, making its distribution inconsistent with that of the observed DWD observations. The OB models' p-values are also lower than those from the IB models and are only greater $0.05$ when including WD kicks. However, Model OB also has a smaller number of DWD systems, so their statistical results are less robust. Model IB+WD kicks give the largest p-value of any model, suggesting that it is the closest to the observed distribution.

As discussed in Section \ref{sec:methodology}, the OB and IB models effectively test different methods of early formation of triples. Although they yield a somewhat similar separation distribution
(see Figure \ref{fig:ICs}), the OB model assumes that the outer binary formed first, and then the inner binary's orbit was limited by stable configurations allowed by the tertiary. In contrast, the IB model assumed that the configurations of the inner and outer binary are independent under the constraint of dynamical stability. As a result of the different assumptions, the OB models generally contain tighter orbits within the triples for both the inner and outer binary (Figure \ref{fig:ICs}). The distributions in Figure \ref{fig:DWD_sep_CDF} and p-values in Table \ref{table: DWD} make it clear that the IB model is far more consistent with the DWD separations in \citet{Heintz22} than the OB models. These results suggest that the formation of the inner and outer binary in triples was independent at birth. This corresponds to an independent fragmentation scenario or dynamical capture during stellar formation. 
Independent hierarchical formation in multiple systems has been supported observationally as well through multiplicity studies \citep[e.g.,][]{Tokovinin14a, Tokovinin14b}. As a result of the initial sampling, the OB modeling contains more triples with close-by tertiaries. From stability, this also leads to tighter allowed orbits within the inner binary. Considering that the observed DWDs in Figure \ref{fig:DWD_sep_CDF} were originally triples, then we would expect the binaries to have smaller separations ($s\lsim10^3$~au), which is not observed. The abundance of close-in PMBs produced by the OB model, therefore, renders it inconsistent with the observations (bottom panel), which favor the IB model.

Although both IB models (with and without kicks) exhibit KS test p-values greater than $0.05$, the model that includes kicks has a much larger p-value. The significance of this difference, taken at face value, can provide support for WD kicks. WD kicks widen the separations of DWDs and unbind the widest ones. From the top panel of Figure \ref{fig:DWD_sep_CDF}, we find that the kicks caused the CDF to have a slight peak around $s=250$~au, which made the rest of the CDF more linear to match observations. 

Overall, the IB triple models reproduce the separation distribution of anomalous DWDs significantly better than 
isolated binary formation and the OB models. This result is statistically robust at $95\%$ confidence because both IB models (with and without WD kicks) have KS test p-values much greater than $0.05$ while the p-value for isolated binary evolution is incredibly small $10^{-13}$. The triple formation channel not only reproduces the anomalous cooling ages of the observed DWDs (Section \ref{subsubsec:DWDs_sep}) but also reproduces their separations better than isolated binary models. Triple evolution and an agnostic kick mechanism at WD formation provide further evidence for a merger fraction of $21-36\%$ amongst wide DWD binaries in the field and a high triple fraction.

\section{Stellar Mergers and Blue Stragglers with Companions}\label{subsec:BSSs}

\subsection{Stellar Mergers and Blue Stragglers Observations}\label{subsubsec:BSSs_obs}

Stellar mergers are common, especially in dense environments \citep{Leonard89}. One-third of all high-mass MS stars may be merger products \citep{deMink14}. In some cases, mergers manifest as peculiar astrophysical transients. For example, mergers containing WDs may manifest as classical novae \citep{Bode08, Starrfield16}. Mergers containing MS stars can instead be observed as luminous red novae (RNe) \citep{Soker06,Tylenda06,Ivanova13b,Pejcha16,MacLeod17, Metzger17,Matsumoto22}. RNe are a class of transients characterized by their large energy output ($10^{45}-10^{47}$~erg), which is just below that of Type Ia supernovae \citep{Bond03,Kulkarni07}. One example of an observed merger is the case of V1309 Sco \citep{Tylenda11}. \citet{Tylenda11}, which was the first LRNe to be observed before, during, and after its final moments. V1309 Sco started as a contact binary with a $1.4$~day period, and over time, its period decreased until the two stars merged and the RNe formed \citep{Tylenda11}. 

Blue Straggler stars (BSSs) are stars found to be bluer or brighter than the main sequence turnoff. BSSs are most often identified in clusters that have known ages derived from their color-magnitude diagrams. BSSs are predicted to be the result of mass transfer or a collision between two MS stars \citep[e.g.,][]{Lombardi95, Lombardi96, Leonard89, Sills97,Sills01,Sills02,Sills05, Freitag05, Ivanova08, Perets09, Gosnell14, Naoz2014}. BSSs can also be the result of the interaction or merger between an MS star and an evolved companion \citep{McCrea64,Chen08,Mathieu09, Gosnell19, Nine20}. Another class of BSSs with similar formation histories are `Blue Lurkers' (BLs), which are identified by their anomalously fast rotation rates \citep{Leiner19}. Their rapid rotation makes them strong candidates for a prior merger event \citep{Sills01, Sills05, Carney05}. The rotation also extends their lifetime as a BSS due to the rotational mixing \citep[e.g.,][]{Sills05, Glebbeek08}. Other observations of peculiar stars include yellow straggler stars \citep{Strom71}, which lie between the end of the MS and the RGB (hypothesized to be evolved BSSs), and red straggler stars \citep{Geller17} which are redder than the RGB. We do not study these in detail here, though our discussed mechanisms may apply to their formation as well.

Observations of star clusters have revealed that BSSs are 2-4 times more likely to contain a companion relative to regular MS stars \citep[e.g.,][]{Mathieu09, Geller12, Nine20}. In the NGC 188 and M67, which have some of the most well-studied BSS populations, the binary fraction of BSSs is of $\sim 76\% \pm 19\%$ and  $\sim 79\% \pm 24\%$, respectively \citep[e.g.,][]{Mathieu09, Geller12, Geller15}. The companions to BSSs are observed at periods ranging from $1$ to $10,000$ days and eccentricities ranging from $0$ to $\sim0.9$ \citep[e.g.,][]{Mathieu09,Geller12, Geller15}. The observed eccentric orbits disfavor isolated binary formation channels since the mass transfer would have presumably circularized the orbits, or they require an additional eccentricity pumping mechanism. If mass transfer between an MS star and an AGB donor did not cause the progenitors to merge, the post-Common Envelope binary could manifest as a blue straggler binary with a hot C/O WD companion \citep{Kippenhahn67, Lauterborn70, Paczynski71, Zenati19}. Such BSSs with hot white dwarf companions have been observed, especially as of recent \citep[e.g.,][]{Gosnell14, Gosnell19, Sun21,Jadhav21,Pandey21, Vaidya22, Nine23}.

The UV-bright WD companions observed nearby in BSSs are a potential marker for a mass transfer history and not a stellar merger. Also, the fact that most BSSs have nearby companions ($<1000$~days) may also lend support to them being the outcome of mass transfer. However, the binary mass transfer channel has a few difficulties. Firstly, observations of some BSSs with low-mass WD companions show that the stars were expected to have undergone unstable mass transfer based on their observed characteristics \citep[e.g.,][]{Pandey21}. Secondly, many BSS binaries are eccentric. These difficulties could be resolved with the three-body channel.
If BSSs are formed through the merger channel, their observed companion would be the tertiary star from the triple. In this case, the observations are actually of the outer binary, which would explain the high eccentricities and the formation of a WD companion in a closer orbit. Namely, the detected WDs could be unresolved inner orbits \citep{Pandey21}. In a triple formation history, the inner binary experiences Kozai-Lidov cycles, which oscillate the eccentricity and inclination of the inner binary \citep{Naoz2016}. During high eccentricity excursions, the pericenter distance decreases, making the probability of stellar interaction high. Collisions induced by destabilized triples can also contribute to their formation, albeit with lower rates \citep{Toonen2022}.

Previous studies have investigated the formation of BSSs from hierarchical triples, \citep[e.g.,][]{Perets09, Naoz2014}, but neglected single and binary stellar evolution. Here, we invoke our simulations of three-body systems--which include triple dynamics, single stellar evolution (with \texttt{SSE}), and detailed binary interactions (with \texttt{COSMIC})--to study BSSs. Specifically, we seek to understand whether the merger channel can still support a fraction of observed BSSs. 
In Section \ref{subsubsec:BSSs_obs_to_sim} we apply our simulated triples that experienced an MS+MS or RG+MS merger to investigate the formation and future evolution of BSSs with companions. We compare these theoretical systems to observations of BSS binaries in the field from \citet{Carney05}, BSS binaries in M67 from \citet{Latham07}, BSS binaries in NGC 188 from \citet{Geller09}, and Blue Lurkers in M67 from \citet{Leiner19}.

\begin{figure*}[h]
\centering
\includegraphics[width=0.65
\textwidth]{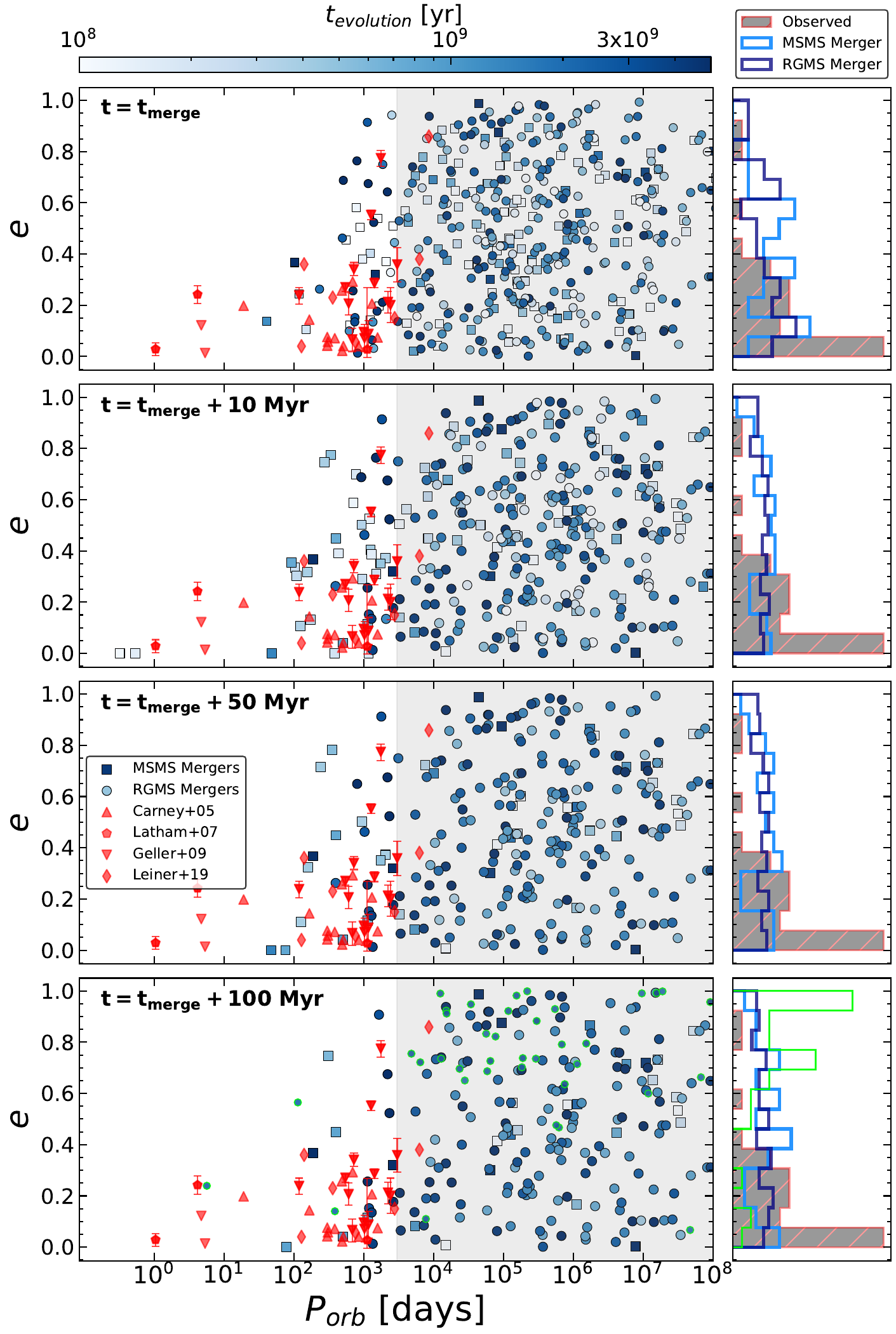}
\caption{Period-Eccentricity diagram of triple systems that experienced an MSMS or RGMS inner binary that merged at different points after the merger time. After the inner binary merges, the triple becomes a `Post-Merger Binary.' Here, we plot the period-eccentricity diagram for this new binary at the time of the merger ($\bf{top~panel}$), 10 million years after the merger ($\bf{second~panel}$), 50 million years after the merger ($\bf{third~panel}$), and 100 million years after the merger ($\bf{fourth~panel}$). If the merger product evolves into a WD, we remove it from the subsequent panels. In red, we overplot observations of Blue Straggler binaries in the field from \citet{Carney05}, Blue Straggler binaries in M67 from \citet{Latham07}, Blue Straggler binaries in NGC 188 from \citet{Geller09}, and Blue Lurkers in M67 from \citet{Leiner19}. In the first panel only, we color the points by the evolution time at the moment that they merged, $t_{merge}$. To the right of every scatter plot, we show the probability density distribution of the simulated MSMS mergers (blue) and RGMS mergers (purple) and compare them to all of the observations (grey-hatched distribution). The shaded region is undetectable by the WOCS RV survey \citet{Geller21}.
}\label{fig:BSS_mergers_after_panel}
\end{figure*}

\subsection{Comparing Observed Blue Stragglers to Theoretical Simulations }\label{subsubsec:BSSs_obs_to_sim}
\subsubsection{Formation from MS+MS and RG+MS Mergers }\label{subsubsubsec:BSS_mergers_explain}
To identify BSS binaries from our simulations, we take the systems where the inner binary began to transfer mass between two MS stars or one MS and one RG star (which includes stars on the AGB). We then follow the detailed mass transfer evolution in \texttt{COSMIC} and only consider those that merge entirely, leaving one MS remnant star behind. We interpret this rejuvenated star as a Blue Straggler (or Blue Lurker) star. 
The triple has now become a binary, where the newly formed primary star is the rejuvenated merger product of two MS stars or an MS+RG, and the secondary is the original tertiary star from the old triple system (often an MS star). We seek to understand how this BSS binary evolves over different timescales after the merger. At the time of the merger, there are $491$ binaries: $237$ from the OB model and $254$ from the IB model. In total, $63\%$ were RGMS mergers, and the rest were MSMS. Most ($65\%$) of RGMS mergers merged in their progenitor triple after $1$~Gyr of three-body evolution, while this is only the case for $16\%$ of MSMS mergers. However, $79\%$ of MSMS mergers occurred after $100$~Myr of triple evolution, and $34\%$ occurred after $500$~Myr. In total, $91\%$ ($446/491$) of the MSMS/RGMS inner binaries merged after $100$~Myr with $47\%$ ($228/491$) merging after $1$~Gyr of three-body evolution. Previous studies of BSS formation from triples found that inner binaries will merge and produce a BSS very rapidly \citep[$\lsim 10^7$~yrs;][]{Perets09,Naoz2014}. These models struggle to reproduce the old BSS populations in clusters \citep[e.g.,][]{Geller15} and even the field BSSs \citep[e.g.,][]{Carney01} because the stars would merge too early and their product would not be observable as a BSS today. With our inclusion of stellar evolution and detailed mass transfer evolution, most merger events occur much later in the evolution of the triple, making the three-body formation channel viable.

After the inner binary merged and the BSS presumably formed, we evolve the rejuvenated BSS binaries for different amounts of time after the time of merger ($t_{merge}$). In Figure \ref{fig:BSS_mergers_after_panel}, we show how the period and eccentricity change for these binaries after $10$~Myr, $50$~Myr, and $100$~Myr of post-merger binary evolution using \texttt{COSMIC}
\footnote{Note that \texttt{COSMIC} assumed efficient tidal evolution which rapidly e=0 for tight binaries \citep{Breivik20}, which underestimates the eccentricity for these systems}. Here, we show the systems that remained on the main-sequence after the merger and discard those that become WDs. At $100$~Myr post-merger, $22\%$ of the merger products have become WDs (see Appendix \ref{app:BSS_WDs}).
The maximum BSS lifetime is assumed to be around $1$-$2$~Gyr \citep{Sills09}. From $100$~Myr to $1$~Gyr post-merger, once tides have settled, the binary orbits do not change significantly, and most BSSs ($80\%$) become WDs. To understand how the MSMS and MSRG merger systems evolve for longer timescales, including post-WD formation, see Figure \ref{fig:BSS_panel_1Gyr} in Appendix \ref{app:BSS_WDs}. 


In Figure \ref{fig:BSS_mergers_after_panel}, the squares represent MSMS mergers, and the circles correspond to RGMS mergers. The color of the points corresponds to the evolution time. In the top panel, we show the orbital structure at the instant that the inner binary completely merged within the triple: the time that the triple became a BSS binary. The circular points often have a darker shade than the squares because the RG+MS binaries generally merge later than the MS+MS mergers. To the right of each period-eccentricity diagram, we show the eccentricity probability density for the MS+MS merger (light blue) and RG+MS mergers (navy blue). Furthermore, we plot the points for observed BSSs and BSS-like systems in black with a red outline. All of the black points are observed systems that are likely to be binaries, with one component being the remnant of an MS+MS or RG+MS collision event. The triangles are the observations of BSS binaries in the field \citep{Carney05}, the pentagons are BSS binaries in M67 from \citet{Latham07} and the diamonds are Blue Lurker binaries in M67 \citep[$4$~Gyr, turnoff mass $1.3$~M$_\odot$][]{Balaguer07} from \citet{Leiner19}. The black circles are BSS binaries in NGC 188 \citep[$7$~Gyr, turnoff mass $1.1$~M$_\odot$][]{Sarajedini99} from \citep{Geller09, Mathieu09, Geller12}. The eccentricity distribution for the observed BSS binaries is also shown (gray hatched distribution). 

In the bottom panel, the green points are the simulated systems that included the impact of WD kicks for each WD companion formed during the $t=100$~Myr post-merger evolution. The adjacent lime histogram shows the distribution of systems with WD kicks. In general, the stellar type of the secondary companions from the observations is unknown, so the fraction of WDs in the observational sample is also unknown. However, the secondary mass distribution of BSS binaries with $\sim 1000$~day periods in NGC 188, as determined through kinematic fitting, peaks at $\sim0.5$~M$_\odot$ \citep{Geller11}. This may suggest that WD companions are common. In Appendix \ref{app:BSS_masses}, we show the secondary mass distributions of our simulated BSS binaries with $P_{orb} < 5000$~days compared to BSS binaries in NGC 188 \citep{Geller11} with similar periods. In general, we find peaks around $0.5 - 0.75$~M$_\odot$ for the BSS companion, consistent with \citet{Geller11}. In our BSS binaries with $P_{orb} < 5000$~days, the companion is a low-mass MS star $87\%$ of the time and a WD $13\%$ of the time $10$~Myr post-merger. After $100$~Myr post-merger, $25\%$ of the BSSs have WD companions.

Since WDs are abundant in BSS systems, we include the systems that experienced small recoil kicks after WD formation to show how kicks affect their eccentricities and periods. As seen from the bottom panel of Figure \ref{fig:BSS_mergers_after_panel}, the binaries with WD kicks achieve wider orbits with higher eccentricities ($e\sim0.8$). Also, most of the close-in systems ($P_{orb}\lsim10^4$~days) reach wider orbits, and most of the wide orbits become unbound. Nearly all of the observations of BSS binaries were taken using the WOCS radial velocity survey, which has a maximum period detection limit of $10^4$~days \citep{Geller12, Geller21} with most being under $3000$~days. Despite this bias, we plot all of our systems over a wide period to give an example distribution for the parent population of BSS binaries.

\subsubsection{Origin of the Orbital Structure of BSS Binaries}\label{subsubsubsec:wd_recoil_kicks}

At $t_{merge}$, the three-body merger channel can successfully reproduce the orbital architectures of most systems, except the shortest BSS binaries. Notably, observed systems with periods above $10$~days have periods and eccentricity that bode well with having previously been hierarchical triples. At this period range, the observed BSS binaries in NGC 188 \citep{Geller09} have a roughly uniform eccentricity distribution, similar to what is expected from the merger channel. On the other hand, the combined observations have a slight preference for smaller eccentricities ($e<0.5$) at this period range. These systems are closely similar to the RGMS mergers with similar periods of $\sim1000$~days but also are consistent with being post-Common Envelope binaries at this period range \citep{Yamaguchi24b}. The preference for shorter eccentricities at these periods may also be a marker for a mass-transfer history within an isolated binary. Among the observed BSS binaries with $P_{orb}~<~1000$~days, most have non-zero, moderate eccentricities. This is surprising considering that they would have presumably circularized rapidly during their mass transfer evolution. A similar phenomenon is observed in post-Common Envelope binaries with similar periods \citep{Yamaguchi24b}. For BSSs, this challenges the mass transfer formation channels and may indicate that another mechanism is pumping eccentricities. 

One eccentricity pumping mechanism to account for the tight BSS binaries with moderate eccentricities is WD formation kicks.
When including the impact of WD kicks, we find that one of our short-period ($P_{orb}<P_{circ}$) BSS binaries with a WD companion escaped its circular orbit and became eccentric upon WD formation (bottom panel of Figure \ref{fig:BSS_mergers_after_panel}). 
Since the orbit is consistent with observations, kicks from WD companions can provide a method of explaining the eccentricities of tight BSS binaries. If this is true, we would predict that BSS binaries with orbital periods less than the circularization period and with non-zero eccentricities have WD companions. This scenario is already supported by observations from calculations of secondary masses to BSSs \citep{Geller11} and observations of hot WDs nearby BSSs \citep[e.g.,][]{Gosnell14, Gosnell19, Sun21,Jadhav21, Vaidya22, Nine23}.



The eccentricity could also be attributed to a distant tertiary that is exciting the eccentricity of the system or cluster dynamics. The tertiary star in a hierarchical triple can still excite eccentricities within inner binaries that have periods around $10$~days if the inner binary contains a WD. For example, \citet{Shariat23} shows that non-interacting WD+MS inner binaries with separations of $0.1$~au ($P_{orb}\sim 10$~days) contain a wide range of eccentricities when they are in hierarchical triples. Moreover, the companions to these binaries have separations between $10$ and $10^4$~au. Therefore, some BSSs binaries may also contain a tertiary as far as $10^4$~au away. Note that the currently observed companion could also be the tertiary.


From \texttt{COSMIC}, an MS+MS merger generally manifests as one inflated MS star, which would have highly efficient tides and an assumed zero eccentricity. Notably, some of the MSMS merger systems sparked a short period of mass transfer during their inflated state or later on as the tertiary evolved off the main sequence. The latter scenario is the same as shown on the right side of Figure \ref{fig:Single_triple_merger}. Such a scenario is an example of triple (or tertiary) mass transfer, which has been noted previously \citep[][]{Gao23,Nine24,Dorozsmai24}, and we find here that it can be a fascinating channel to form close-in BSS binaries.




A significant fraction of BSS stars have observed Barium enrichment \citep{Nine24}, which is a signature of AGB mass transfer. However, if most BSSs are formed from AGB mass transfer, BSS binaries with $P_{orb}\gsim1000$~days are expected to be rare since unstable mass transfer would lead to a merger or a tighter orbit. If we consider that the Ba enrichment is from an AGB+MS merger, which is the scenario for most of the RG+MS mergers, these orbital periods would be attainable. We discuss the AGB merger scenario more below.

As mentioned, a distinct sign that a star transferred mass or merged with an AGB companion is the enrichment of \textit{s}-process elements such as Barium (Ba) \citep{Iben83,Busso99}. \citet{Nine24} found that $43\pm11\%$ of the BSSs that they observed in three different clusters, including M67, are Ba enriched. Many Ba-enriched stars have observed companions \citep{Escorza19, Jorissen19, Nine24}, most of which are WDs. Most Ba binaries have orbital parameters consistent with AGB mass transfer, though some exhibit puzzling configurations. Namely, a handful of `unexplained' Ba-enriched binaries have been discovered \citep{Escorza19, Jorissen19} at periods that cannot be reproduced with binary evolution modes. Three of these binaries have short periods ($P_{orb}<700$~days) but modest eccentricities \citep{Escorza19}, and two have a long period ($P_{orb}\sim10^4$~days) and very high eccentricity ($e\gsim0.9$) \citep{Escorza19,Jorissen19}. If we consider that the Ba enrichment is from an AGB+MS merger, which most of our RG+MS mergers are, then these orbital periods can be abundantly formed with the triple merger channel (Figure \ref{fig:BSS_mergers_after_panel}, top panel). As discussed earlier, the unexplained short-period Ba-enriched binaries could have a similar formation to the close-in, eccentric BSS binaries that we describe above. Also, since these Ba stars have WD companions, kicks could play a role to slightly alter eccentricities. The long-period `unexplained' binaries are occupied in the eccentricity-period diagram in Figure \ref{fig:BSS_mergers_after_panel}, especially by RG+MS mergers in triples in the top and bottom panels. Previous studies also support that a significant fraction of Ba stars may have formed from hierarchical triples \citep{Gao23}.

Overall, we find that the triple+merger channel can reproduce the periods, eccentricities, and many other discussed characteristics of most observed BSS binaries. However, observations of BSS binaries clearly show that $\sim75$-$80\%$ of them have companions with orbital periods less than $3000$~days \citep{Geller11, Geller15}. In the triple+merger channel, most tertiaries lie at periods above this range and beyond the RV detection limit of the BSS surveys, making them observable as single BSSs. Therefore, for BSS binaries with periods below $3000$~days, the stable mass transfer channel is likely to be dominant. Nevertheless, there exists a handful of observed BSS binaries with peculiar properties in this period regime that can be naturally explained by the triple+merger channel or by accounting for additional binary evolution mechanisms. Furthermore, based on our models, it is likely that the $20$-$25\%$ of BSSs without nearby companions formed via the triple+merger channel and have tertiary stars beyond the detection limit. Previous studies have estimated that the fraction of BSSs formed in triples is closer to $10\%$ \citep{Antonini16}. Lastly, it is important to note that mass transfer will not always lead to circular orbits, which is a feature of {\tt COSMIC} models \citep[e.g.,][]{HamersMT}. 

\section{WD+MS and WD+RG Merged Inner Binaries}\label{sec:WD_MSRG}

\begin{figure}
\centering
\includegraphics[width=1.0
\columnwidth]{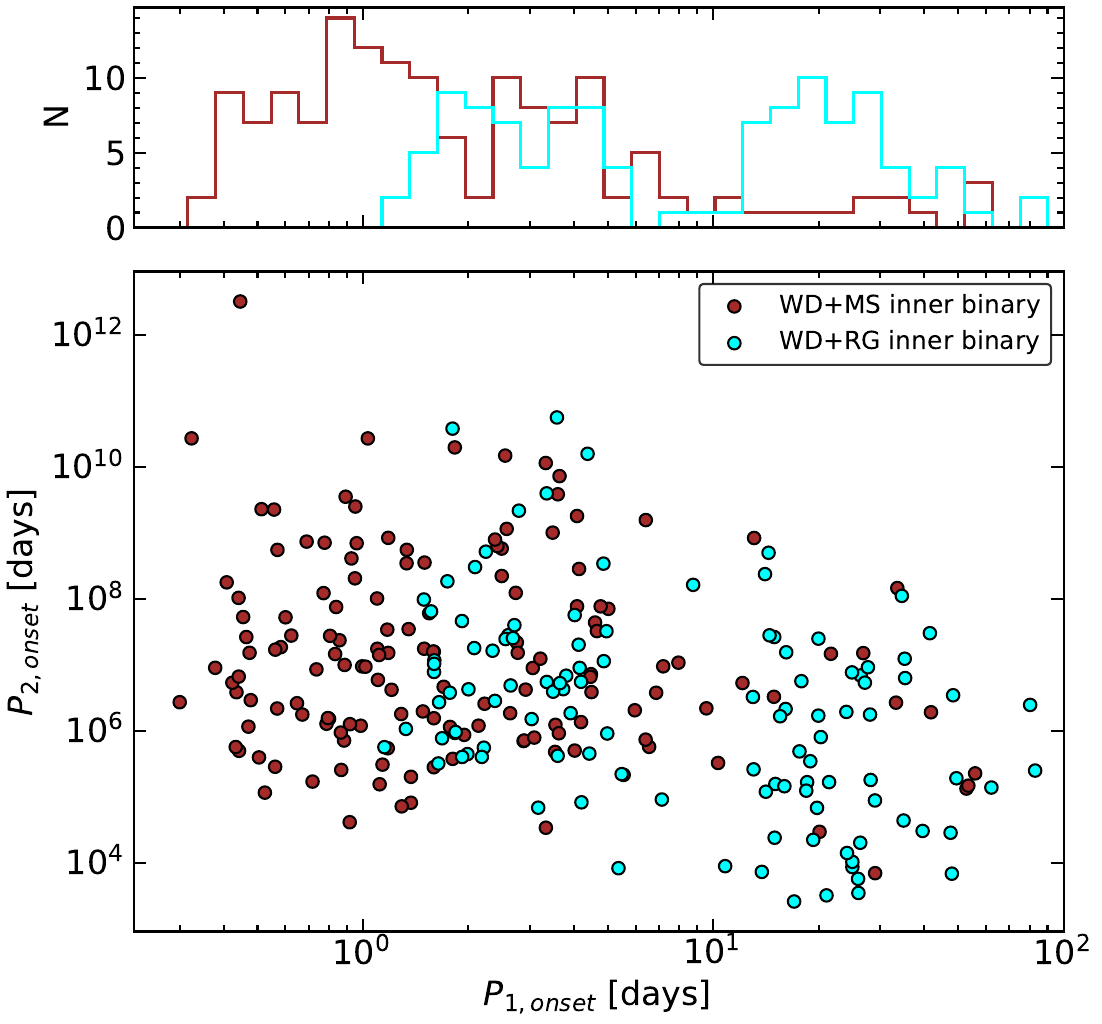}
\caption{ The inner and outer period in the triples with inner merged WD+MS (red) or WD+RG (cyan) binaries at the onset of WD mass transfer ($P_{1,onset}$ and $P_{2,onset}$, respectively). On the top, we plot the unnormalized distribution of $P_{1,onset}$ for both types of inner binaries. 
}\label{fig:CV_like_orbits}
\end{figure}

A quarter of all merged triples include mergers between a WD+MS or WD+RG (Table \ref{tab:merger_types_models}). Before the WD formed, many of these inner binaries already experienced a common envelope phase. For most of the WD+RG mergers, the first mass-transfer episode stopped when the WD formed, and the second mass-transfer episode (which led to the merger) began as the secondary reached the first RGB or AGB. 
Here, we are interested in the triple's orbital configuration at the onset of mass transfer with the WD primary. In Figure \ref{fig:CV_like_orbits}, we plot the parameter space for the orbital periods in the triple at the onset of WD mass transfer. We denote the inner (outer) period at the onset of WD mass transfer as $P_{1,onset}$ ($P_{2,onset}$). In the WD+MS mergers, we also include companions that are in the Hertzsprung Gap. 

In general, the RG stars begin mass transfer at larger $P_{1,onset}$ since they have larger radii. At the onset of WD mass transfer, the inner periods are bimodal for both WD+MS and WD+RG systems. This bimodality is especially pronounced for the WD+RG accreting binaries and can provide clues to their formation history. All of the short-period WD+RG binaries ($P_{1,onset}<10$~days) have C/O WDs, whereas most of the longer-period ones have He WD primaries. The WD type gives insight into how the WD progenitor evolved, and when it began its first (pre-WD) common envelope evolution. For example, most of the He WDs progenitors began mass transfer earlier than the C/O WD progenitors, meaning their envelopes were stripped earlier from the secondary. Furthermore, C/O WDs generally have smaller radii than He WDs, so they only initiate mass transfer with larger, more evolved companions.

All of these mass transferring WD+MS and WD+RG in this figure eventually merged. During the pre-merger mass transfer, the WD+MS binaries would be observed as a CV-like system with a gradually decreasing period. For low-mass WDs, WD+MS merged systems are expected to be a common outcome from their unstable common envelope evolution. This merger scenario could potentially account for the abundance of high WD masses and the lack of He WD accretors in CVs \citep{Schreiber16,Belloni18}. Namely, the lower mass WDs would have led to a quick merger and failed to produce a CV. From the abundance of high-mass WDs in CVs, it is therefore expected that a population of WD+MS merger products, or `failed CVs,' exist in the Galaxy. These merger remnants are expected to look like a modified RG star with a dense core and an inflated envelope generated from the MS component \citep{Rui24}. In fact, most of the WD+MS merger products from our simulations lie on the RGB, according to \texttt{COSMIC}. These unusual RG stars may exhibit distinct photometric, asteroseismic, and surface abundance signatures \citep{Rui21,Li22,Deheuvels22,Matteuzzi23, Rui24}.

\section{Other Outcomes}\label{sec:other_outcomes}

\subsection{Double Mergers}\label{subsec:double_mergers}

After the inner binary merges in a hierarchical triple, a PMB is formed, and in $10\%$ of all simulations, the PMB also experiences mass transfer evolution. One example of common envelope evolution in a PMB is displayed in Figure \ref{fig:Single_triple_merger}. Interestingly, in $4.4\%$ ($36/810$) of all PMBs, the PMB underwent unstable mass transfer and merged again, leaving behind a single star. These systems began as hierarchical triples; the inner binary merged, and then the remaining binary merged, leaving {\it a single} star behind.
This `double merger' scenario occurred $36$ times in our simulations and possess a diverse array of formation histories and potential future evolution. Among the double mergers, the first inner binary merger was between an MS+MS, RG+MS, and WD+MS in $26$ ($72\%$), $6$ ($27\%$), and $4$ ($11\%$) cases, respectively. 
For all $4$ WDMS inner binary mergers (which left behind a single WD), the second merger was also a WD+MS merger. Since this post-merger, WD was already massive ($\gsim 0.8$~M$_\odot$), in all $4$ cases, the second merger event ignited a super-Chandrasekhar explosion, which did not leave behind any remnant star. The commonality of such an explosion from the double merger channel is unknown, though observations suggest that many local massive WDs may be merger products \citep{Hallakoun23}. Moreover, we show earlier (Section \ref{subsec:DWDs}) that many field WDs in wide DWDs experienced a prior merger (Section \ref{subsec:DWDs}). With a considerable number of post-merger massive WDs from triples, the double merger channel may be a notable method of producing single-degenerate Type Ia supernovae in the Galaxy. In observations, double mergers will likely be difficult to distinguish from single mergers. Similar to single mergers, double mergers may contain photometric, asteroseismic, and surface abundance features \citep[e.g.,][]{Rui21}, though future work is required to disentangle them from regular stars and single mergers.

\subsection{Neutron Stars from Mergers}\label{subsubsec:neutron_stars}

Another interesting, yet small ($2\%, 18/810$), subset of our merged systems is those that produced a Neutron Star (NS) in the PMB after $10$~Gyr of total evolution time. This includes an NS+MS, NS+RG, and NS+WD binaries. Initially, the masses from the Kroupa IMF were constrained to be below $8$~M$_\odot$ \citep{Shariat23}, so producing an NS requires either an MS progenitor greater than $8-10$~M$_\odot$ (which must have formed from a previous merger or mass transfer) or an accretion-induced core-collapse of a WD. $9$ of the binaries did not remain unbound after the neutron star supernova kick. $2$ of them remained bound but later merged because of their high post-kick eccentricity, leaving behind one NS. Of the $7$ surviving NS binaries ($6$ with MS companions and $1$ with a WD companion), one formed from accretion-induced WD collapse, and the rest became massive stars after previous episodes of mass transfer or a previous merger. The separations of the NS+MS binaries from this channel range from $100$ to $3000$~au, while the $2$ NS mergers were short-lived at $\lsim1$~au eccentricities. 

In general, the NS in NS+MS binaries formed after two $\sim 5$~M$_\odot$ MS stars in the inner binary coalesced with minimal mass loss to produce a rejuvenated $\sim 10$~M$_\odot$ star on the MS, which later evolved into a NS. Previous simulations support that little mass is lost during MS+MS mergers \citep{Lombardi02, Dale06, Glebbeek08}. Before WD formation, this rejuvenated massive star may exhibit a strong, large-scale magnetic field \citep[e.g.,][]{Donati09,Ferrario15,Fossati15,Schneider16} that hints at its merger origin.

The NS+WD PMB formed as the result of a WD further collapsing into an NS after mass transfer with a donor. In the original triple, two MS stars merged to create one $7$~M$_\odot$ MS star in a binary with a $1$~M$_\odot$ MS star. As the primary MS star evolved, mass transfer began, and it eventually became a $1.3$~M$_\odot$ Ne/O WD. Then, as the secondary evolved, the two stars began common envelope evolution, which led to the WD collapsing into an NS. The NS is also a pulsar rotating  $400$ times per second while the secondary is a $0.1$~M$_\odot$ He WD. The supernova induced a $40$~km/s kick, making the final orbital period of the NS+WD binary only $100$~minutes. During the evolution of this binary, it may have been observed as a CV (during the MS accretion onto the WD), and at its final configuration, would possess signatures as a millisecond pulsar with strong gravitational wave emission from its nearby WD companion \citep[observed examples in the ATNF pulsar catalog,][]{Manchester05}. Eventually, gravitational waves will extract energy from the binary, which will shrink the orbit and eventually put the two stars in contact. In this binary, the WD mass is small enough ($\lsim0.2$~M$_\odot$) such that the NS+WD mass transfer will be stable \citep{Bobrick17}, making the system an ultra-compact X-ray binary \citep[UCXB;][]{vanHaaften12}.

NS binaries were not the primary focus of this study. However, their formation from solar-type triples presents an intriguing scenario for creating close, intermediate, and wide NS binaries. Intermediate-separation NS binaries have been observed with {\it Gaia} \citep[e.g.,][]{Andrews22, EB24a, EB24b}. Moreover, the triple channel proved to form close NS binaries with MS and WD companions, which can be observed as symbiotic XRBs \citep[e.g.,][]{Hinkle06, Yungelson19, Hinkle19, Hinkle24, Nagarajan24} during their evolution. Future work that focuses on larger initial masses can examine the efficiency of the triple merger channel in forming NS binaries.

\section{Conclusions}\label{sec:conclusions}
Over the past decades, observations of stars have challenged our understanding of single stellar evolution and often required the consideration of binary and higher-order systems to explain their formation. Fortunately, most stars form and evolve with one or multiple stellar companions \citep{Raghavan2010, Moe17}, making stellar interactions highly probable. Previously, \citet{Shariat23} evolved solar-type triples for over $10$~Gyr using detailed dynamical simulations with stellar evolution. In their simulations, more than half of all solar-type triples experience inner binaries that merge within $12.5$~Gyr. In this study, we focus on investigating the formation and evolution of stellar triples with merged inner binaries containing main-sequence, red giant, and white dwarf stars. Specifically, we explored the outcomes of merged MS+MS, RG+MS, WD+MS, and WD+RG merger inner binaries.

On top of the dynamical simulations, we track the detailed mass transfer history in these triples using the \texttt{COSMIC} binary stellar evolution code. This includes the first episode of mass transfer, which led to a merged inner binary, and the subsequent evolution of the post-merger binary. Often, we evolve the post-merger binary for Myr to Gyr after the merger event. We then compare our results to the observation of peculiar stellar systems to examine the following question: does a post-merger binary (PMB) retain any signature of its triple past?

Throughout this paper, we discuss the different potential outcomes of PMBs and compare them to observations of wide DWDs and BSS binaries to test the three-body formation channel. Our main results are summarized as follows:

\begin{enumerate}

\item \textit{Triple formation channel for DWD binaries:} We find strong evidence that DWD binaries with discrepant ages \citep{Heintz22,Heintz24} are the result of previous triples that experienced a merger. In particular, we show that the triple merger channel reproduces the discrepant ages (Figure \ref{fig:DWD_cool_mass}) and the separation distribution of observed DWDs, whereas isolated binary evolution cannot (Figure \ref{fig:DWD_sep_CDF} and Table \ref{table: DWD}). The model that matched most closely with the observations (IB+kicks), which assumes that an agnostic mechanism causes a small $\sim0.75$~km/s kick in the system. Presumably, kicks are a result of asymmetric mass loss during WD formation. The kicks fundamentally shift the separations of the simulated binaries to create a greater statistical match. This provides further support for the existence of WD recoil kicks.

\item \textit{Formation hierarchy in stellar triples:} Our triple models that assumed independent orbital periods between the inner and outer binary are consistent with observations of wide DWDs, while our other model is not (Figure \ref{fig:DWD_sep_CDF} and Table \ref{table: DWD}). The consistency between the IB model and observations gives insight into the order of star formation during the early phases of multiple stellar evolution. Namely, it supports a scenario where the inner binary and the tertiary form independently and at roughly the same time.

\item \textit{Significant merger fraction in wide DWD binaries:} The strong consistency of the triple channel with anomalous DWDs suggests that $26$-$54\%$ \citep{Heintz22} of wide DWDs were previously three-body systems, and today, the more-massive WD is a merger product. Another fraction of DWDs previously had a companion that was made unbound due to flyby stars, and WD kicks over $10$~Gyr of evolution \citep[e.g.,][]{Shariat23}. $61\%$ of the simulated triples from \citet{Shariat23} that formed a DWD became unbound from the tertiary, and out of all the initial triples, $14\%$ became isolated DWD binaries with $s>100$~au. Assuming that $30\%$ of all solar-type stars ($<8$M$_\odot$) were born in triples \citep{Shariat23}, then $4\%$ of all solar-type stars become isolated DWDs after unbinding with a tertiary. In total, we estimate that $44\pm14\%$ of local DWDs were birthed in triple-star systems and are now isolated binaries because of a previous merger or because they became unbound with a prior companion.

\item \textit{Formation channel for BSS binaries:} The orbital configuration of PMBs that had MSMS or RGMS mergers closely matches with most ($P_{\rm orb}>10$~days) observed BSS binaries (Figure \ref{fig:BSS_mergers_after_panel}). Based on this, we predict that the $20$-$25\%$ of BSSs without nearby companions formed via stellar collisions and have very long period ($P_{orb}>3000$~days) companions. However, current observations show that most observed BSSs in clusters have close companions \citep[$P_{orb}<3000$~days;][]{Geller11, Geller15}, which supports that a majority of BSSs formed via stable mass transfer. Still, the eccentricities in some of the close BSS binaries are inconsistent with stable mass transfer, but can naturally be explained by triple dynamics or by considering an additional eccentricity pumping mechanism (e.g., WD kicks). We also show that collisions in hierarchical triples often occur after Myr to Gyr of triple evolution, producing BSS ages consistent with old clusters. Lastly, similar to the mass transfer channel, the triple+merger channel would also produce enriched chemical abundances, fast rotation rates, and companion masses observed in many BSSs. 

\item \textit{Signatures of WD Formation Kicks:} We find imprints of WD recoil kicks consistently throughout our analyses. In the case of DWDs, WD kicks serve to qualitatively change the modality and shape of the separation distributions. They create a bimodal distribution, leading to a greater statistical consistency between theory and observations. In the case of BSS binaries, WD kicks lead to an explanation for the eccentric short-period binaries observed, which would, therefore, presumably have a WD companion. The eccentricity can also be attributed to a distant, unresolved tertiary (see Section \ref{subsubsec:BSSs_obs_to_sim}) or cluster dynamics.

\item \textit{WDMS and WDRG mergers:} We find that $19.8\%$ ($7.1\%$) of PMBs formed after a WDMS (WDRG) merged inner binary. Figure \ref{fig:CV_like_orbits} depicts the inner and outer periods of hierarchical triples with merged WDMS and WDRG inner binaries. The inner periods are plotted at the first moment of WD mass transfer (Figure \ref{fig:CV_like_orbits}), which initiates an unstable interaction and leads to their eventual merger. These mergers are expected to produce CV-like systems and transient novae (see Shariat et al. in prep for more details).

\item \textit{Other Outcomes:} We show that triples with mergers can also provide a formation channel for close, intermediate, and wide NS+MS or NS+RG binaries. In this channel, the NS either formed from an MS progenitor that became a massive star after a merger or from accretion-induced WD collapse. We also explore triples that experience two mergers ($5\%$ of PMBs), which provide a method of producing single-degenerate Type Ia supernova. Understanding the dynamical origins and identifying formation markers of merged stars is increasingly important, especially as methods of distinguishing merger remnants from regular stars continue to advance.
\end{enumerate}

\section{acknowledgements}\label{acknowledgements}
We thank Tyler M. Heintz and J.J. Hermes for bringing our attention to the age-discrepant DWDs in the Gaia sample. We thank the anonymous referee for a constructive report. This work used computational and storage services associated with the Hoffman2 Shared Cluster provided by UCLA Office of Advanced Research Computing’s Research Technology Group.  S.N. acknowledges the partial support from NASA ATP 80NSSC20K0505 and from the NSF-AST 2206428 grant, as well as thanks Howard and Astrid Preston for their generous support.

\appendix 
\section{The long-term evolution of MS+MS and RG+MS mergers}\label{app:BSS_WDs}
After $100$~Myr after an RG+MS merger, most of the merger products have evolved to WDs. In this case, the merger product is too evolved to be a bright star on the MS, so it will not be observable as a BSS. Here, we show the orbits of these evolved PMBs as the merger product evolves to a WD. In most cases, the PMB circularized as the progenitor of the WD, which is the merged star, evolved off the main sequence. The post-merger evolution is dependent on \texttt{COSMIC} models. For details on the evolution of BSSs that formed through stellar mergers, see Section \ref{subsec:BSSs}.

\begin{figure}[!htp]
\centering
\includegraphics[width=0.5\columnwidth]{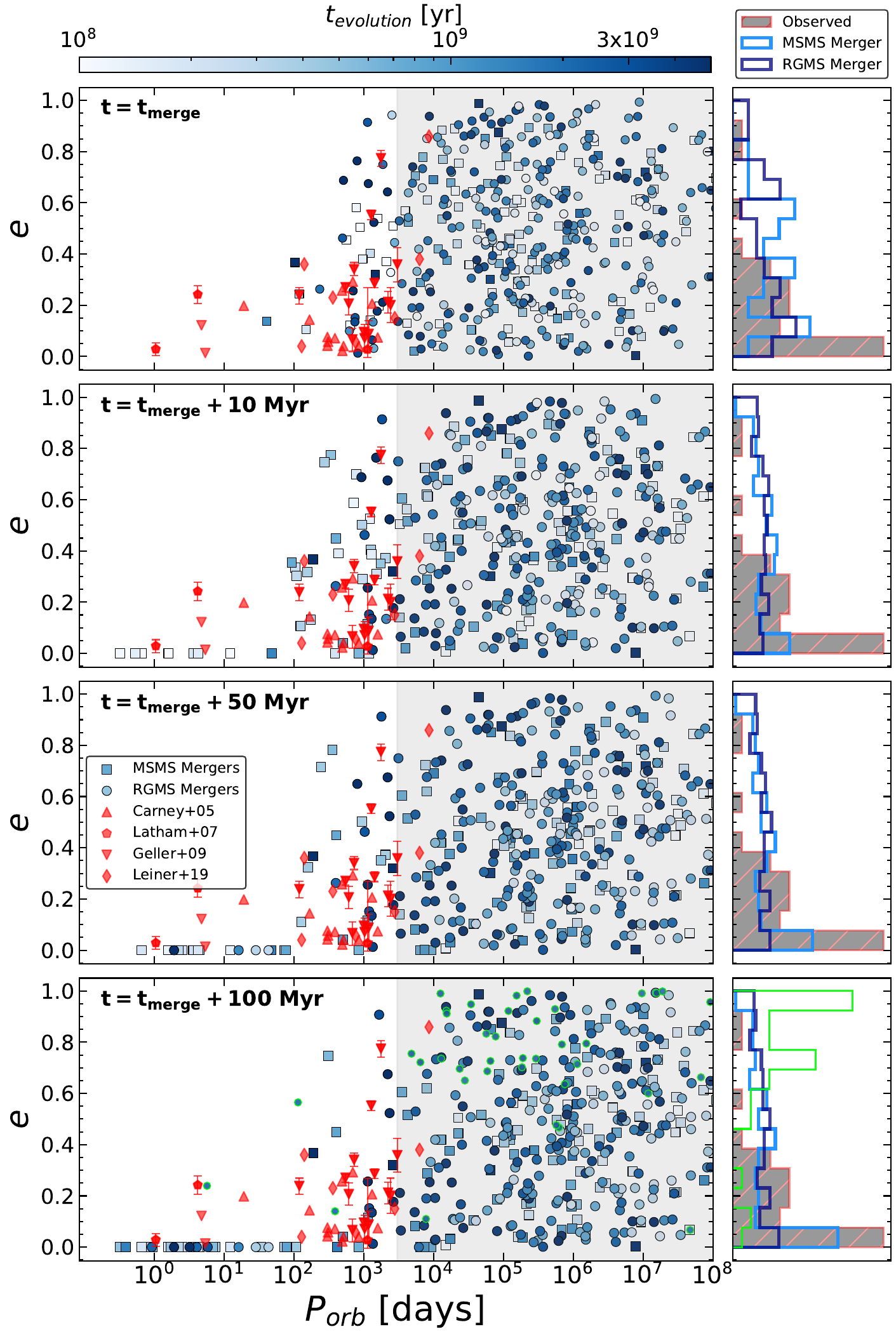}
\caption{Period-Eccentricity diagram of triple systems that experienced an MS+MS or RG+MS inner binary that merged at different points after the merger time. Same as Figure \ref{fig:BSS_mergers_after_panel} but also includes merged stars after they evolved to become WDs.
}\label{fig:BSS_panel_1Gyr}
\end{figure}
 
\section{Companion masses in BSS binaries}\label{app:BSS_masses}
The companions to BSSs can give insight into the formation history of the BSS. \citet{Geller11} found that the BSS binaries with periods around $1000$~days have companion masses ($M_2$) that peak around $\sim0.5$~M$_\odot$. 
These masses were determined by fitting the mass function statistically. Here, we show the distribution for the companion masses for the simulated BSS binaries with similar periods from the triple-merger channel. We plot the $M_2$ distribution for our BSS binaries $10$~Myr, $50$~Myr, and $100$~Myr after the merger event. Similar to \citet{Geller11}, the $M_2$ distribution peaks around $0.4$-$0.8$~M$_\odot$. As the binary evolves, the distribution becomes more localized in the aforementioned mass range.

\begin{figure}[!htp]
\centering
\includegraphics[width=0.45\columnwidth]{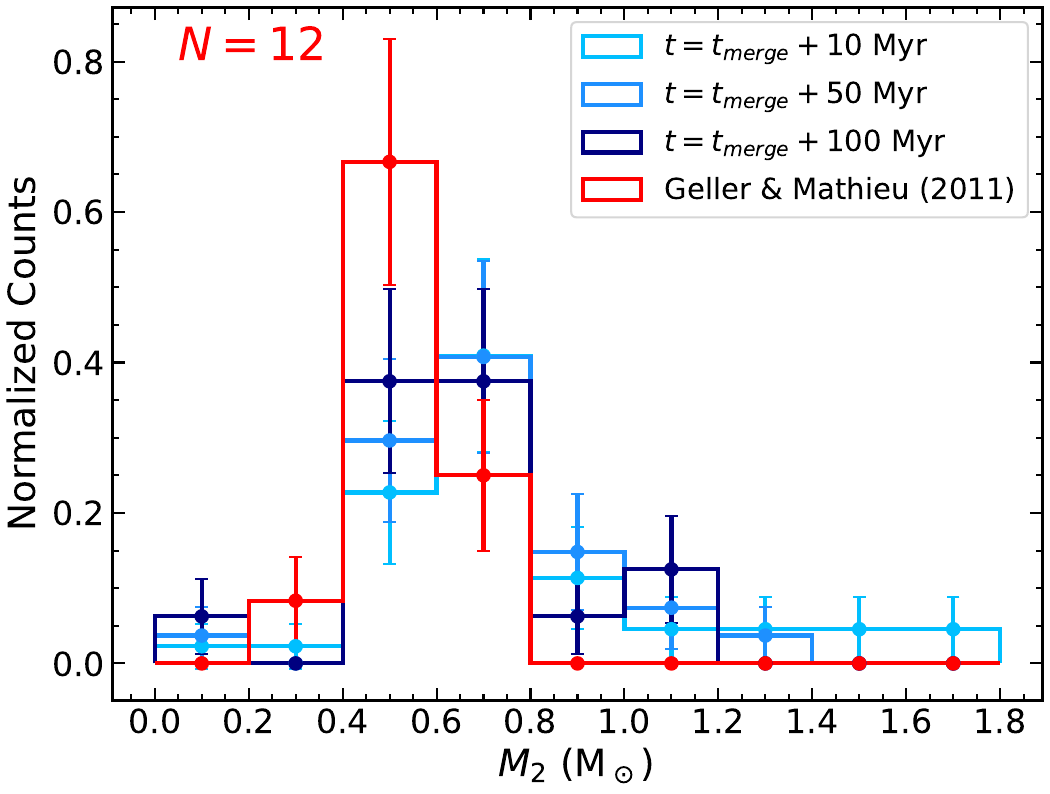}
\caption{Masses of BSS binaries with orbital periods of order $1000$~days. The red distribution represents the companion mass of $12$ BSS binaries in NGC 188 with periods of order $1000$~days. To compare with the observed distribution, we plot the companion for the simulated BSS binaries formed from the triple merger channel. We only consider the simulated systems with $P_{orb}<5000$~days at $10$, $50$, and $100$~Myr after the merger.
}\label{fig:BSS_masses}
\end{figure} 
\clearpage
\bibliography{references}
\end{document}